\newif\iftechreport%
\techreporttrue%

\RequirePackage{etoolbox}
\ifdef{\iftechreport}{}{\newif\iftechreport}

\documentclass[acmsmall]{acmart}

\setcopyright{cc}
\setcctype{by}
\acmJournal{PACMMOD}
\acmYear{2026} \acmVolume{4} \acmNumber{1 (SIGMOD)} \acmArticle{43} \acmMonth{2} \acmPrice{}\acmDOI{10.1145/3786657}

\usepackage{multirow}

\usepackage{color} 
\usepackage{cuted}
\usepackage{fancyhdr}

\usepackage{graphicx}
\usepackage{caption}
\usepackage{subcaption}
\usepackage{amsthm}
\usepackage{thmtools,thm-restate}
\usepackage[ruled,vlined,linesnumbered]{algorithm2e}
\SetKwProg{Fn}{Function}{}{}
\SetVlineSkip{0pt}
\SetAlCapNameFnt{\small}
\SetAlCapFnt{\small}
\SetAlFnt{\small}
\makeatletter

\patchcmd\algocf@Vline{\vrule}{\vrule \kern-0.4pt}{}{}
\patchcmd\algocf@Vsline{\vrule}{\vrule \kern-0.4pt}{}{}
\patchcmd{\@algocf@start}{-1.5em}{0pt}{}{}
\makeatother

\usepackage{multicol}
\usepackage{multirow}
\usepackage{booktabs}
\usepackage[dvipsnames]{xcolor} 

\usepackage[capitalise,noabbrev]{cleveref}

\newcommand{\rv}[1]{#1}

\definecolor{myBrown}{RGB}{128,23,31}
\definecolor{myGreen}{RGB}{29, 131, 72}

\newcommand{\rone}[1]{{\color{black} #1}}
\newcommand{\rtwo}[1]{{\color{black} #1}}
\newcommand{\rthree}[1]{{\color{black} #1}}
\newcommand{\opttree}{\textrm{HIRE}}

\usepackage[linewidth=1pt]{mdframed}
\usepackage{hyperref}

\mdfdefinestyle{@commentstyle}{
  backgroundcolor=black!15,
  hidealllines=true,
  innertopmargin=\topskip,
  innerbottommargin=\topskip,
  skipabove=0pt,
  skipbelow=0pt,
}

\newmdenv[style=@commentstyle]{reviewbox}

\apptocmd\normalsize{%
   \setlength{\abovedisplayskip}{2pt}
   \setlength{\belowdisplayskip}{2pt}
   \setlength{\abovedisplayshortskip}{2pt}
   \setlength{\belowdisplayshortskip}{2pt}
}{}{}

\usepackage[inline]{enumitem}
\setlist{noitemsep,partopsep=0pt,topsep=0pt}
\setlist[itemize,1]{leftmargin=2.2em}

\makeatletter
\algocf@newcmdside@kobe{Broker}{%
    \KwSty{Broker:}%
    \ifArgumentEmpty{#1}\relax{ #1}%
    \algocf@block{#2}{end}{#3}%
    \par
}

\algocf@newcmdside@kobe{Provider}{%
    \KwSty{Data Provider:}%
    \ifArgumentEmpty{#1}\relax{ #1}%
    \algocf@block{#2}{end}{#3}%
    \par
}

\usepackage{fancyhdr}

\makeatletter
\fancypagestyle{revisionstyle}{%
  \fancyhf{}
}
\makeatother

\begin{document}


\title{\opttree{}: A Hybrid Learned Index for Robust and Efficient Performance under Mixed Workloads}


\author{Xinyi Zhang}
\affiliation{%
  \institution{Hong Kong Baptist University}
  \country{Hong Kong SAR}
}
\email{csxyzhang@comp.hkbu.edu.hk}

\author{Liang Liang}
\affiliation{%
  \institution{EPFL}
  \country{Switzerland}
}
\email{liang.liang@epfl.ch}

\author{Anastasia Ailamaki}
\affiliation{%
  \institution{EPFL}
  \country{Switzerland}
}
\email{anastasia.ailamaki@epfl.ch}

\author{Jianliang Xu}
\affiliation{%
  \institution{Hong Kong Baptist University}
  \country{Hong Kong SAR}
}
\email{xujl@comp.hkbu.edu.hk}



\renewcommand{\shortauthors}{Xinyi Zhang, Liang Liang, Anastasia Ailamaki, and Jianliang Xu}
\begin{abstract}

Indexes are critical for efficient data retrieval and updates in modern databases. Recent advances in machine learning have led to the development of learned indexes, which model the cumulative distribution function of data to predict search positions and accelerate query processing. While learned indexes substantially outperform traditional structures for point lookups, they often suffer from high tail latency, suboptimal range query performance, and inconsistent effectiveness across diverse workloads. To address these challenges,  this paper proposes 
\opttree{}, a hybrid in-memory index structure designed to deliver efficient performance consistently. 
\opttree{} combines the structural and performance robustness of traditional indexes with the predictive power of model-based prediction to reduce search overhead while maintaining worst-case stability. Specifically, it employs (1) hybrid leaf nodes adaptive to varying data distributions and workloads, (2) model-accelerated internal nodes augmented by log-based updates for efficient updates, (3) a non-blocking, cost-driven recalibration mechanism for dynamic data, and (4) an inter-level optimized bulk-loading algorithm accounting for leaf and internal-node errors. Experimental results on multiple real-world datasets demonstrate that \opttree{} 
outperforms both state-of-the-art learned indexes and traditional structures in range-query throughput, tail latency, and overall stability. Compared to state-of-the-art learned indexes and traditional indexes, HIRE achieves up to $41.7\times$ higher throughput under mixed workloads, reduces tail latency by up to 98\% across varying scenarios. 

\end{abstract}

\maketitle

\sloppy
\newtheorem{assumption}[theorem]{Assumption}

    
\section{Introduction} \label{sec:intro}

Indexes are essential components of modern DBMSs, and their efficiency impacts overall database performance for both retrieval and update operations~\cite{silberschatz2011database}. The rapid advancement of machine learning has driven the emergence of learned indexes, which integrate predictive models with traditional indexing techniques~\cite{kraska2018case}.

Learned indexes employ approximation models to learn the cumulative distribution function (CDF) of data, predicting data locations rather than traversing fixed algorithmic paths. 
Early studies demonstrated exceptional search performance, suggesting their potential to replace traditional index structures~\cite{kipf2020radixspline, stoian2021towards, hadian2021shift}. However, maintaining updatability remains a significant challenge. Dynamic data updates can induce model drift, degrading prediction accuracy or necessitating computationally expensive retraining, thereby reducing the efficiency of learned indexes.

To address these challenges, updatable learned indexes, such as FIT-ting Tree~\cite{galakatos2019fiting}, ALEX~\cite{ding2020alex}, PGM~\cite{ferragina2020pgm}, and LIPP~\cite{wu2021updatable}, have been proposed, accompanied by extensive analyses~\cite{ferragina2020learned, liu2024learned, sun2023learned, wongkham2022updatable, liu2025good}, optimizations~\cite{zhang2024revisiting, zhang2024making, choi2024can, chockchowwat2022tuning, wang2025new, zhong2022learned}, and applications~\cite{chesetti2024evaluating, schmidt2024predicate, chatterjee2024limousine, liang2024swix, lan2024fully, yang2023flirt}.
However, while several studies suggest that updatable learned indexes are ``almost ready'' for separated update and query workloads~\cite{wongkham2022updatable}, 
our experimental results reveal that learned indexes still exhibit high tail latency and inefficient range query performance under mixed workloads.
%
%
%
As shown in Figure~\ref{fig:moti}, a comparison of state-of-the-art learned indexes with B+-tree under a mixed workload highlights three key limitations: (1) \textit{inconsistent performance} across different workloads; (2) \textit{suboptimal performance for range queries}; 
and (3) \textit{unstable query/update latency}. A detailed discussion of these limitations is provided in Section~\ref{sec:background_motivation}.

The primary reason for these limitations is that \textit{updatable learned index designs often prioritize optimal performance for point queries under specific conditions, at the cost of worst-case inefficiency and instability.} For example, PGM \cite{ferragina2020pgm} employs an index-level buffer to mitigate model drift during insertions, which renders range queries highly inefficient due to the need to search multiple trees. Similarly, LIPP \cite{wu2021updatable} sacrifices data locality and balanced structure  to enhance model accuracy, achieving low-cost point lookups and insertions but compromising range query performance. 

In this paper, we address the performance inefficiency and instability of 
updatable learned indexes to achieve robust and efficient performance under mixed workloads involving range queries.
%
To this end, we propose \opttree{}, a hybrid in-memory index that integrates the stability of traditional indexes, such as B+-tree, with the performance advantages of learned indexes that leverage data CDFs to accelerate searches. 

Specifically, \opttree{} employs a balanced-tree framework to ensure consistent performance across diverse workloads. At the leaf layer, we design two types of leaf nodes to maintain robust performance under varying conditions. At the internal layers, we introduce model-accelerated internal nodes, augmented by a log-based update mechanism that reduces frequent movements of keys and child node pointers. This design enhances query speed while minimizing update overhead. Furthermore, to mitigate high tail latency during retraining, \opttree{} adopts a non-blocking, cost-driven approach that combines multi-threaded concurrent operations with recalibration. 
For index construction, we propose an inter-level optimized bulk-loading algorithm that optimizes errors in multiple layers, enabling better post-construction performance. Compared to state-of-the-art updatable learned indexes and traditional indexes,  \opttree{} demonstrates superior performance across diverse mixed workloads.

Our main contributions are summarized as follows:
\begin{itemize} 
\item We analyze state-of-the-art updatable learned indexes and identify their trade-offs through experiments.
Our analysis reveals that current designs often result in poor data locality and unbalanced structures, which degrade range query performance and increase tail latency versus traditional indexes.
\item We propose \opttree{}, a novel in-memory index to address the performance inefficiency and instability of updatable learned indexes under mixed workloads involving range queries. To ensure high query efficiency while optimizing update and tail latency, \opttree{} incorporates several key components: a hybrid leaf node structure, model-accelerated internal nodes with log-based updates,  a non-blocking, cost-driven  recalibration algorithm, and a bulk-loading construction process optimized for low error. 
\item We implement a prototype of \opttree{} and conduct extensive experiments on multiple datasets. Compared to state-of-the-art learned indexes and traditional indexes, \opttree{} achieves significant performance improvements in range query efficiency and latency stability. Specifically, \opttree{} delivers up to $41.7\times$ higher throughput on mixed workloads, while reducing tail latency by up to 98\%. Moreover, \opttree{} consistently demonstrates robust performance across all evaluated scenarios.
\end{itemize}

The rest of the paper proceeds as follows. We present the background and motivation of our work in \cref{sec:background_motivation}. We introduce our novel  \opttree{} index in \cref{sec:overall_design} and discuss its operations in \cref{sec:hire_operations}. The evaluation results are shown in \cref{sec:experiments}. Finally, we review the related work in \cref{sec:related_works} and conclude the paper in \cref{sec:conclusion}.

\section{Background and Motivation} \label{sec:background_motivation}

\begin{figure*}
    \centering
    \begin{minipage}{0.45\linewidth}
    \centering
        \begin{subfigure}{\linewidth} 
            \centering
            \includegraphics[width=\linewidth]{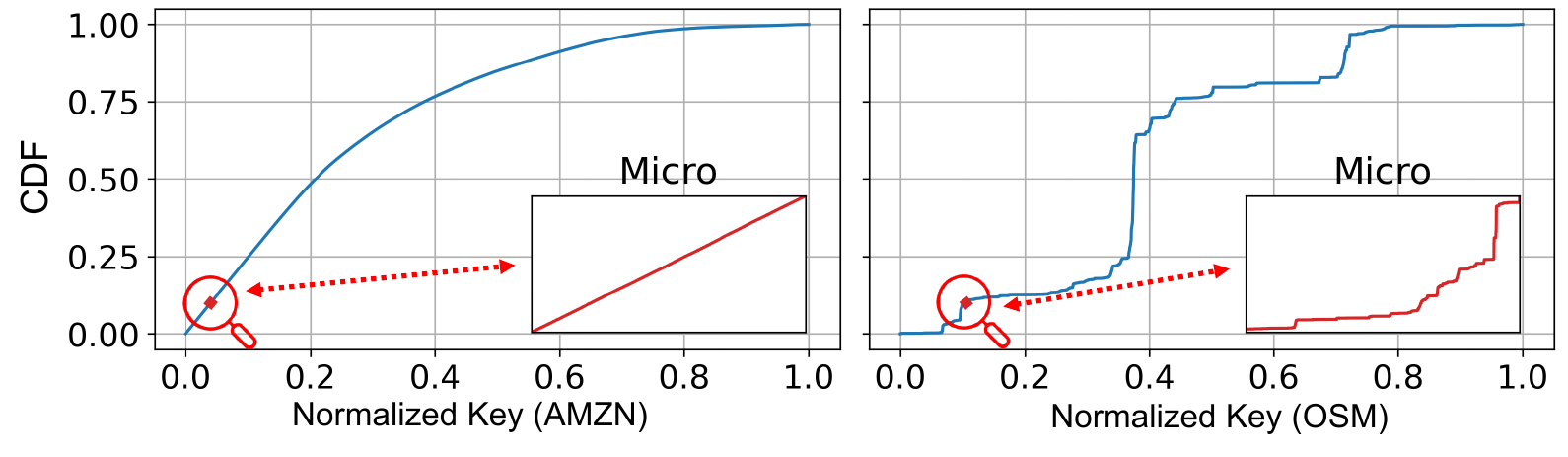}
            \caption{}
            \label{fig:moti_cdf_comp}
        \end{subfigure}
        \iftechreport
        \begin{subfigure}{\linewidth} 
            \centering
            \includegraphics[width=\linewidth]{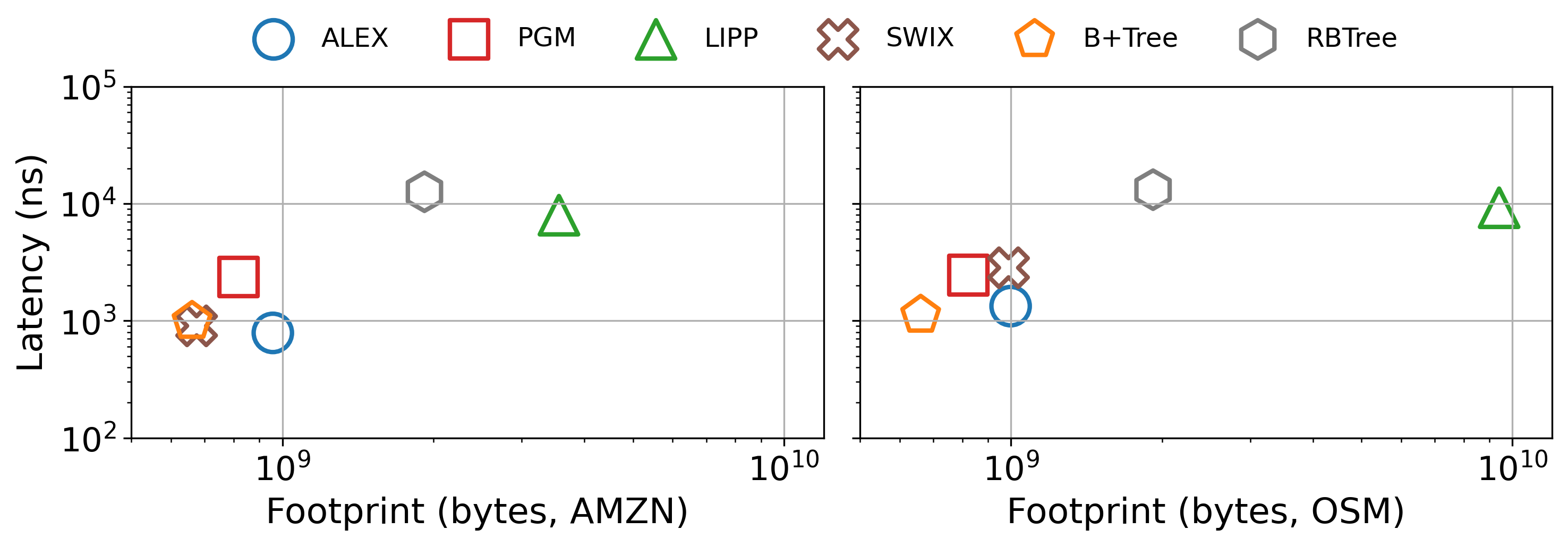}
            \caption{}
            \label{fig:moti_mem_latency}
        \end{subfigure}
        \else
        \begin{subfigure}{\linewidth} 
            \centering
            \includegraphics[width=\linewidth]{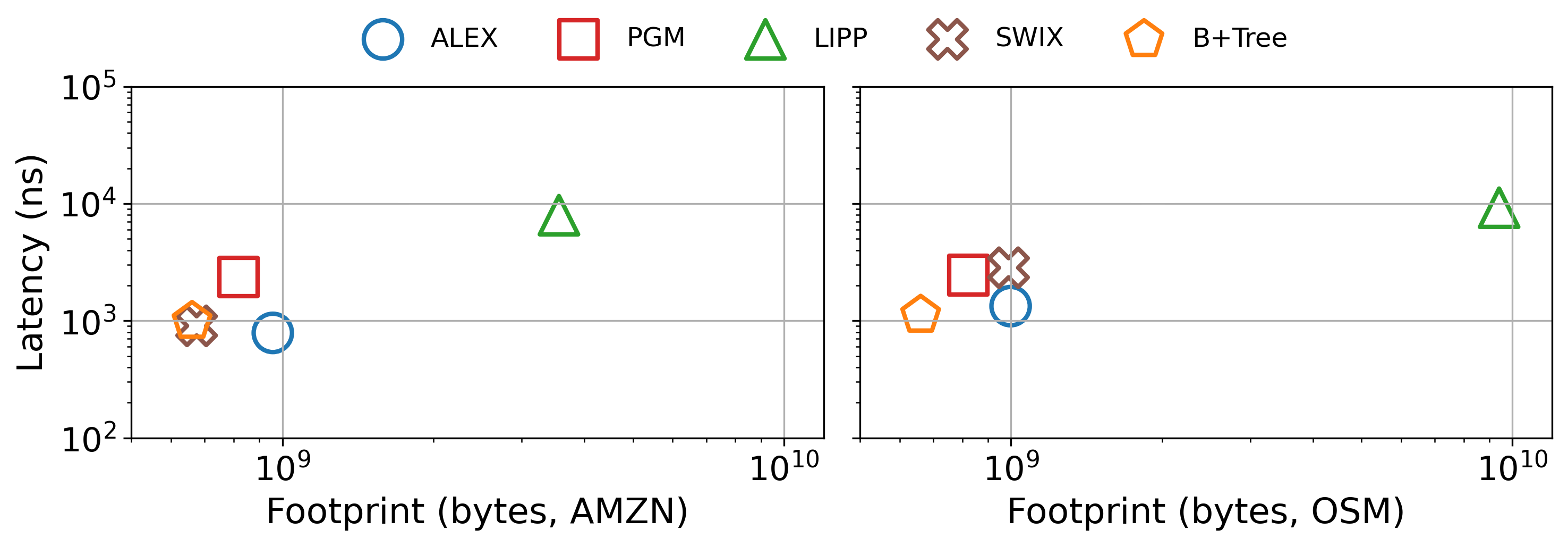}
            \caption{}
            \label{fig:moti_mem_latency}
        \end{subfigure}
        \fi
    \end{minipage}
    \begin{minipage}{0.48\linewidth}
        \begin{subfigure}{\linewidth} 
            \centering
            \includegraphics[width=\linewidth]{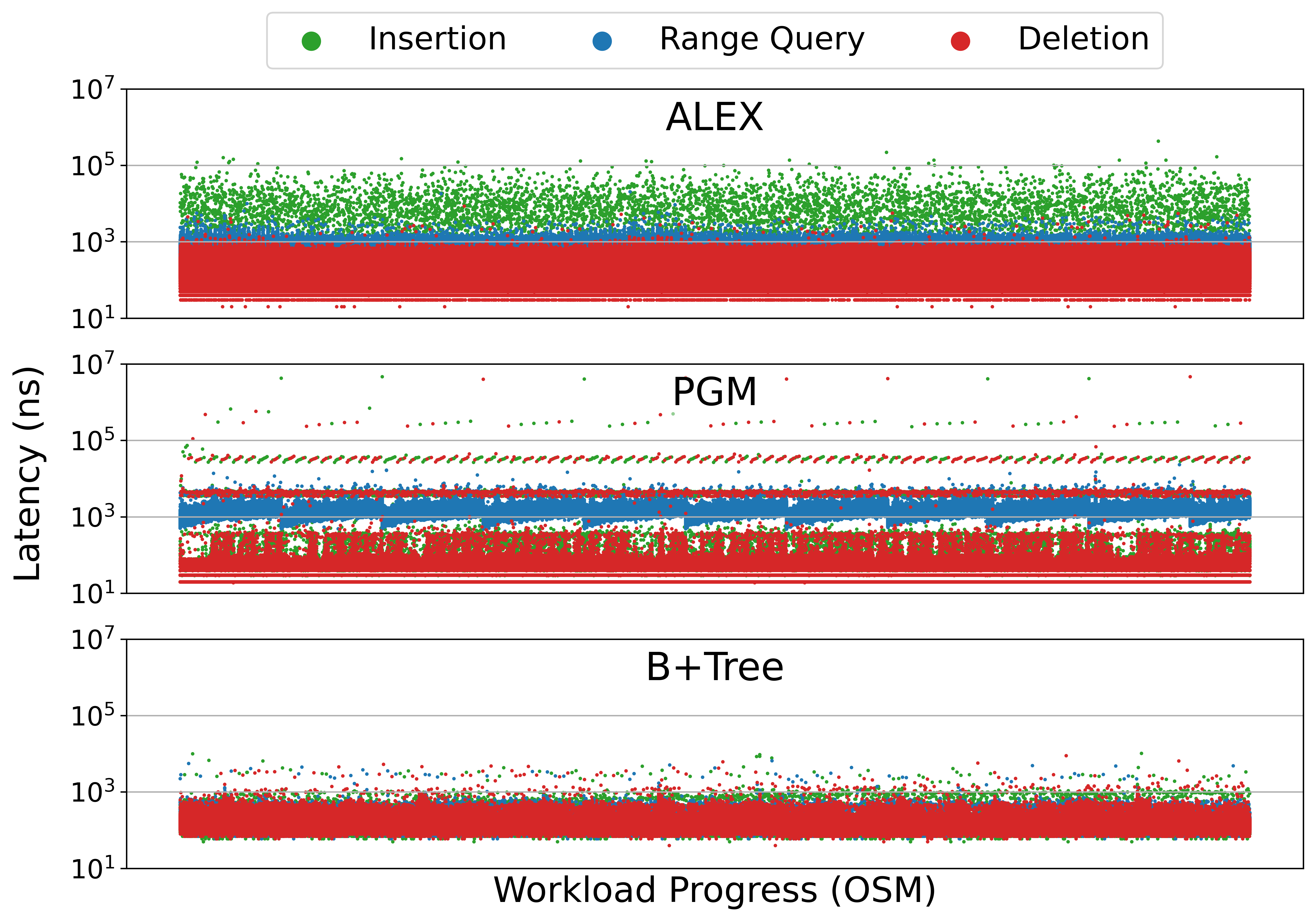}
            \caption{}
            \label{fig:moti_log}
        \end{subfigure}
    \end{minipage}
    \caption{Illustration of performance limitations of existing learned indexes under a \rthree{balanced mixed workload (Query : Insert : Delete = 1:1:1)}. (a): Distributions of two SOSD datasets. (b): Performance comparisons of different indexes. (c): Latencies of insertion, deletion, and range query operations on the \textsf{OSM} dataset.}
    \label{fig:moti}
\end{figure*}

Consider a table $\mathcal{T}$, where each data entry $o_i \in \mathcal{T} $ is represented as a tuple $\langle k_i, v_i \rangle$, with $k_i \in \mathbb{R}$ as the key and $v_i$ as the associated value. To retrieve the entry $o_i$, we search for $k_i$ by locating its position $p_{k_i}$ in the storage structure. This search can be accelerated by indexes. To support efficient range queries, the indexed $k_i$'s are stored in monotonically increasing order (i.e., $\forall k_m < k_n,\ p_{k_m} < p_{k_n}$). Thus, $p_{k_i}$ can be represented as an integer that increases monotonically with $k_i$. The index serves as a mapping between the search key $k_i$ and its storage location $p_{k_i}$. 

In contrast to traditional indexes like B+-tree, learned indexes employ machine learning models to approximate the data distribution and predict key locations~\cite{kraska2018case}. Specifically, a model $F$ learns the cumulative distribution function (CDF) of the indexed keys, enabling it to predict a search bound for a given key. 
Traditional indexes, such as B+-tree, use a parameter‐based search bound, such as the fanout $f$, to guide searches. In contrast, learned indexes rely on an error‐based search bound $\epsilon$, representing the potential deviation between the predicted and actual key location. This deviation requires a ``last‐mile search'' to locate the key precisely. Crucially, the model 
$F$ must be strictly monotonic (i.e., $\forall k_m < k_n,\ F(k_m) < F(k_n)$). Without strict monotonicity, the last‐mile search range could span the entire table $\mathcal{T}$, rather than being bounded by $\epsilon$.

\begin{table*}[]
\tiny
\centering
\begin{tabular}{|c|l|ll|l|l|ll|}
\hline
\multicolumn{1}{|c|}{\multirow{2}{*}{\textbf{Index}}} & \multirow{2}{*}{\textbf{Structure}} & \multicolumn{2}{l|}{\textbf{Range Query}}                                                                                                                                                             & \multirow{2}{*}{\textbf{Insertion}}                                           & \multirow{2}{*}{\textbf{Deletion}} & \multicolumn{2}{l|}{\textbf{Recalibration}}                                                                      \\ \cline{3-4} \cline{7-8} 
\multicolumn{1}{|c|}{}                                &                                     & \multicolumn{1}{l|}{\textit{\textbf{Point Lookup}}}                                                                     & \textit{\textbf{Range Scan}}                                                &                                                                               &                                    & \multicolumn{1}{l|}{\textit{\textbf{Local}}}                                          & \textit{\textbf{Global}} \\ \hline
\textbf{ALEX}~\cite{ding2020alex}                                                  & DAG                                 & \multicolumn{1}{l|}{\begin{tabular}[c]{@{}l@{}}Hierarchical Model \\ with Correction\end{tabular}}                      & \begin{tabular}[c]{@{}l@{}}Sequential Scan \\ with Skip\end{tabular}        & Data-level Buffer                                                             & Mask                               & \multicolumn{1}{l|}{\begin{tabular}[c]{@{}l@{}}Merge, Split, \\ Retrain\end{tabular}} & Top-down                 \\ \hline
\textbf{PGM}~\cite{ferragina2020pgm}                                                   & \multicolumn{1}{l|}{\begin{tabular}[c]{@{}l@{}}Mult. Balanced \\ Tree \end{tabular}}                  & \multicolumn{1}{l|}{\begin{tabular}[c]{@{}l@{}}Hierarchical Model\\ with Correction \\ Tree-buffer Search\end{tabular}} & \begin{tabular}[c]{@{}l@{}}Sequential Scan \\ with Random Scan\end{tabular} & Index-level buffer                                                            & Mask                               & \multicolumn{1}{l|}{\begin{tabular}[c]{@{}l@{}}Merge, Retrain\end{tabular}}          & Bottom-up                \\ \hline
\textbf{LIPP}~\cite{wu2021updatable}                                                  & Imbalanced Tree                      & \multicolumn{1}{l|}{Hierarchical Model}                                                                                 & \begin{tabular}[c]{@{}l@{}}Random Scan with \\ Sequential Scan\end{tabular} & Inplace                                                                       & Mask                               & \multicolumn{1}{l|}{\begin{tabular}[c]{@{}l@{}}Split, Retrain\end{tabular}}          & Top-down                 \\ \hline
\textbf{SWIX}~\cite{liang2024swix}                                                  & Two-level Queue                     & \multicolumn{1}{l|}{\begin{tabular}[c]{@{}l@{}}Two-level Model \\ with Correction\end{tabular}}                         & \begin{tabular}[c]{@{}l@{}}Sequential Scan \\ with Skip\end{tabular}        & \begin{tabular}[c]{@{}l@{}}Data-level Buffer\\ Node-level Buffer\end{tabular} & Mask                               & \multicolumn{1}{l|}{\begin{tabular}[c]{@{}l@{}}Merge, Split, \\ Retrain\end{tabular}} & Bottom-up                \\ \hline
\textbf{B+-tree}~\cite{comer1979ubiquitous}                                               & Balanced Tree                        & \multicolumn{1}{l|}{Hierarchical Search}                                                                                & Sequential Scan                                                             & Inplace                                                                       & Inplace                            & \multicolumn{1}{l|}{\begin{tabular}[c]{@{}l@{}}Merge, Split\end{tabular}}           & Bottom-up                \\ \hline
\iftechreport
\textbf{RBTree}~\cite{guibas1978dichromatic}                                                & Balanced Tree                        & \multicolumn{1}{l|}{Hierarchical Search}                                                                                & Random Scan                                                                 & Inplace                                                                       & Inplace                            & \multicolumn{1}{l|}{\begin{tabular}[c]{@{}l@{}}Merge, Split\end{tabular}}           & Bottom-up                \\ \hline
\fi
\textbf{\rthree{\opttree{}}} & \rthree{Balanced Tree}  & \multicolumn{1}{l|}{\begin{tabular}[c]{@{}l@{}}\rthree{Hierarchical Hybrid} \\ \rthree{Search}\end{tabular}}   & \multicolumn{1}{l|}{\begin{tabular}[c]{@{}l@{}}\rthree{Hybrid Sequential} \\ \rthree{Scan} \end{tabular}} & \multicolumn{1}{l|}{\begin{tabular}[c]{@{}l@{}} \rthree{Data-level Buffer} \\ \rthree{Inplace} \end{tabular}}& \multicolumn{1}{l|}{\begin{tabular}[c]{@{}l@{}}\rthree{Mask} \\ \rthree{Inplace} \end{tabular}}  & \multicolumn{1}{l|}{\begin{tabular}[c]{@{}l@{}}\rthree{Non-blocking} \\ \rthree{Recalibration} \end{tabular}} & \multicolumn{1}{l|}{\begin{tabular}[c]{@{}l@{}}\rthree{Subtree} \\ \rthree{Replacement} \end{tabular}}  \\ \hline
\end{tabular}
\caption{Comparison of Index Designs} 
\label{tab:design_comparison}
\end{table*}

The predictive model is central to the design of learned indexes. For static data, a complex mapping function can eﬀectively map keys to their positions with a small $\epsilon$, achieving near-constant-time lookups. 
However, dynamic operations, such as insertions and deletions, disrupt the sorted key order, causing $\epsilon$ to grow linearly with the number of updates. 
\rtwo{As updates accumulate, the prediction error grows and degrades model accuracy. Therefore, naive designs (e.g., simply using the model in the leaf nodes of the B+-tree) are ineffective unless additional mechanisms are introduced to handle updates.} Retraining the mapping function is computationally expensive, making it unsuitable for dynamic updates. Consequently, updatable learned indexes typically employ simpler piecewise linear models to construct a hierarchical structure relating keys and their positions, sacriﬁcing some accuracy for faster updates. To mitigate the impact of dynamic operations, these indexes incorporate strategies such as update mitigation techniques, recalibration methods, and structural supports (summarized in Table~\ref{tab:design_comparison}). 
These design choices reflect a trade-off: \textit{learned indexes prioritize fast lookups, often at the expense of degraded worst-case performance under highly dynamic workloads}. We discuss representative designs below: 





\begin{itemize}
\item ALEX~\cite{ding2020alex}, a pioneering updatable learned index, employs linear interpolation to approximate the data distribution. This linearization enhances the generalization of the linear model and reduces the search space to a reasonable $\epsilon$. The interpolated space also accommodates insertions without significantly affecting model accuracy. However, linearization reduces data locality. When the data distribution is highly non‐linear and cannot be approximated by a single linear model within the error bound $\epsilon$, ALEX recursively partitions the data, using multiple linear models. This top‐down partitioning may result in an unbalanced index structure. While ALEX excels in point queries and updates in dynamic settings, its suboptimal data locality and potential for an unbalanced structure impair range query performance and introduce instability in operational latency.

\item PGM~\cite{ferragina2020pgm} 
uses piecewise linear models to segment data and constructs an internal node layer from these segments. However, its update mechanism adopts a semi-dynamic approach,  
buffering updates before periodically merging them into existing segments, which triggers retraining. This two-stage approach requires searching both the buffer and data segments during queries, impairing data locality. Consequently, PGM prioritizes read-heavy workloads, achieving superior average query and update performance in such scenarios. However, this results in higher query and update tail latency in write-intensive environments.

\item LIPP~\cite{wu2021updatable} 
aims for perfect precision by reducing $\epsilon$ to zero. However, it requires top‐down node splitting for non‐linear data distributions, which may result in a highly unbalanced structure. LIPP mitigates the impact of insertions on model accuracy by reserving extra space within nodes. While this enables excellent point query and insertion performance, it significantly compromises range query efficiency and introduces high tail latency.
\end{itemize}

In summary, while existing methods can constrain model error within a predefined threshold, they often compromise index balance and data locality. Balanced structures are crucial for minimizing performance variability and reducing tail latency. Maintaining good data locality is also essential for efficient range queries. 

\rthree{\cref{fig:moti}, following the default settings in \cref{sec:experiments}, validates our analysis using two SOSD datasets: \textsf{AMZN} and \textsf{OSM}.} As shown in Figure~\ref{fig:moti_cdf_comp}, \textsf{AMZN} is learned-index-friendly due to its linear micro-level distribution, 
despite a non-linear macro-level distribution. In contrast, \textsf{OSM}'s non-linearity at both macro and micro levels challenges accurate linear modeling.
%
%
%
Figure \ref{fig:moti_mem_latency} presents a comparison of various indexes on the two datasets under a mixed workload of random insertions, deletions, and range queries (returning $100$ results). 
The lower-left region of the plot represents ideal performance (low latency and low memory overhead), while the upper-right region indicates suboptimal performance.
B+-tree 
performs well in latency and size across both datasets, leveraging its robust design to ensure data locality and structural balance. 
ALEX and SWIX achieve latency comparable to B+-tree on model-friendly distributions (\textsf{AMZN}) but suffer significant degradation on non-linear distributions (\textsf{OSM}).
PGM, while maintaining memory efficiency similar to B+-tree, incurs higher range query latency due to the need to search across multiple index-level buffers. 
\iftechreport
Red-black tree (RBTree), despite  achieving balance through node coloring and rotations, exhibits suboptimal latency for updates and range queries. Its binary structure disperses data, necessitating frequent rotations and pointer updates during modifications and jumps during range queries.
\fi
LIPP's compromise in data locality and structural balance lead to poor performance. 
%
Figure~\ref{fig:moti_log} shows time-varied latencies on the \textsf{OSM} dataset. ALEX and PGM exhibit significant latency fluctuations. For ALEX, varying segment sizes lead to variable latencies during local recalibration, while structural imbalance triggers costly structural recalibration during insertions, particularly at deeper levels. PGM's high tail latency stems primarily from index buffer merging during recalibration, which is necessary to maintain model accuracy and data locality. In contrast, B+-tree's uniform leaf node size, balanced structure, and strong data locality significantly reduce tail latency.

In essence, 
the inconsistent performance of learned indexes is an inherent drawback: while they leverage models for fast searches under ideal conditions, their reliance on data distribution assumptions limits generality. Unlike traditional indexes, which provide robust and balanced performance, learned indexes excel only when the data closely adheres to the learned pattern. Deviations necessitate corrective searches or retraining, resulting in degraded performance for range queries, high tail latency, and increased memory consumption. This motivates a re-examination of learned index design to achieve robustness and stability under mixed workloads involving range queries. Integrating design principles from the proven B+-tree structure offers a promising direction.

\section{\opttree{} Design} \label{sec:overall_design}

Based on observations in \cref{sec:background_motivation}, we identify three key design goals for a general-purpose and updatable learned index: 1) robustness to diverse data distributions without relying on specific assumptions; 2) high efficiency for both point and range queries; and 3) stable latency for operations. 
To achieve these goals, we propose \opttree{}, a hybrid in-memory index that innovatively combines the strengths of B+-tree and learned indexes. As shown in \cref{fig:architecture}, \opttree{} adopts a traditional hierarchical index structure with a hybrid model to adapt to varying data distributions. It employs two node structures: leaf nodes with a compact data layout to accelerate range queries, and internal nodes with model-based acceleration and log-based updates for efficient traversal and update operations. To ensure stable latency, \opttree{} maintains a balanced tree structure. A non-blocking, cost-model-driven recalibration mechanism is designed to prevent latency spikes, ensuring the index remains fully available during maintenance. Additionally, an inter-level optimized bulk-loading algorithm is developed to enhance the initial construction efficiency. Together, these novel designs enable \opttree{} to deliver robust and efficient performance across diverse mixed workloads involving range queries.


\subsection{Leaf Layer with Hybrid Nodes} \label{sub_sec:leaf_layer}

\begin{figure}[tp]
    \centering
    \includegraphics[width=0.8\linewidth]{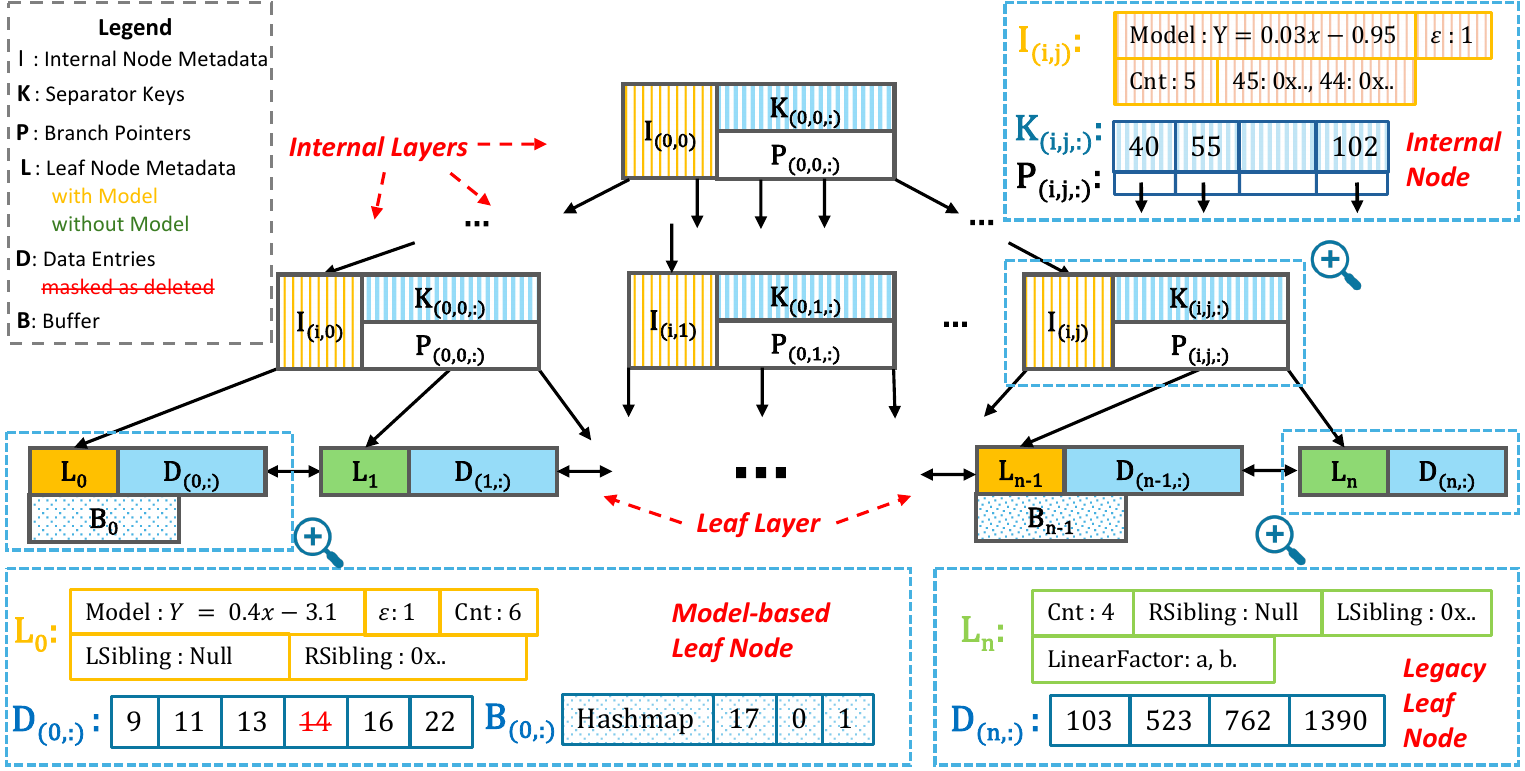 } 
    \caption{Structure of \opttree{}}
    \label{fig:architecture}
\end{figure}

As illustrated in \cref{fig:architecture}, the leaf layer of \opttree{} stores actual data in a linked list composed of compact nodes. We have opted for a compact layout because leaf nodes constitute the majority of nodes in an index.  Alternative approaches, such as gap arrays, while enhancing model accuracy, introduce significant storage overhead and compromise data locality, which degrade range query performance. 

The leaf layer is hybrid, containing two different node types to adapt to various data distributions. For highly linear data segments, model-based leaf nodes employ a linear model within an error bound (e.g., \(\epsilon\)) to approximate key-position relationships. This approach increases the capacity of a leaf node without compromising query performance, thereby reducing the number of nodes, index size, and tree height. To address the challenges of updates in model-based leaf nodes, we design an optimized buffer that supports $O(1)$ operations for queries, insertions, and deletions. This includes a masking scheme for deletions that  preserves model validity without incurring costly data movement. 

For data segments that defy linear approximation, \opttree{} reverts to legacy leaf nodes, which are similar to B+-tree nodes. These legacy leaf nodes feature a smaller capacity, utilize SIMD-based linear search, and support for in-place updates. To unify the design, we incorporate a dynamic optimization that records regression factors in each legacy leaf and leverages a cost model to opportunistically merge adjacent legacy nodes into efficient model-based nodes. 
\subsection{Model-accelerated Internal Nodes} \label{sub_sec:internal_node_design}

Each internal node stores a list of keys from its child nodes along with their corresponding pointers. \rthree{Unlike leaf nodes, which also need to support efficient scans, internal nodes access behave more like point lookups. To optimize internal nodes, we integrate the search efficiency of learned indexes with a log-based update mechanism inspired by LSM-trees~\cite{DBLP:journals/acta/ONeilCGO96}. For rapid lookups, each node employs a linear model over its keys. To improve model precision and update efficiency, the slope of the linear model is adjusted, and keys are remapped to accommodate insertions when child nodes split. Furthermore, by batching updates rather than executing them individually, the log structure serves as a buffer, avoiding the overhead of immediate retraining and delaying model invalidation after new insertions.} To ensure accurate searches, the update log is accessed during each search and is designed to be lightweight and cache-efficient. As will be detailed in \cref{subsub_sec:updates}, our log-based update mechanism minimizes data movement while supporting fast lookups.

\subsection{Non-Blocking Cost-Driven Recalibration}

Updates in \opttree{} can trigger two types of recalibration. While low-cost structural adjustments, such as splits and merges, can be performed immediately, the expensive retraining of the model-based leaf node presents a significant challenge, as a naive approach would block operations and lead to high tail latency. 
%
%
To mitigate high tail latency and operation blocking, we propose a non-blocking, cost-model-driven recalibration mechanism. This mechanism uses background threads to execute costly retraining and structural adjustments without stalling foreground operations. Concurrently, a built-in cost model proactively determines the optimal moment to trigger these tasks by balancing performance gains against their overhead. This approach enables \opttree{} to adapt to changes in data distribution while maintaining stable, low tail latency. The detailed procedures are discussed in \cref{sub_sec:retrain}.

\subsection{Inter-level Optimized Bulk-Loading}

To construct an efficient \opttree{} from scratch using existing data, we propose a bulk-loading algorithm that optimizes segment fitting across the leaf and internal layers to enhance model performance. Starting with sorted key-value pairs, the algorithm constructs an “optimal” leaf layer by using the model-based and legacy leaf nodes described in \cref{sub_sec:leaf_layer}, minimizing the number of leaf nodes while ensuring prediction errors within $\epsilon$. Simultaneously, splitting keys from the leaf layer are used to build upper-level nodes. During this process, the upper level may re-select splitting keys to refine lower-level segmentation, reducing prediction errors across layers. This inter-level optimization enhances the efficiency of model-accelerated internal node searches while maintaining $\epsilon$ at the leaf level. Further details are provided in \cref{sub_sec:bulk_load}.

\section{\opttree{} Operations} \label{sec:hire_operations}


    

\subsection{Queries} \label{sub_sec:point_lookups}
\subsubsection{Point Lookup Query} \label{sub_sub_sec:point_lookups}

\begin{figure}[tp]
    \centering
    \includegraphics[width=0.7\linewidth]{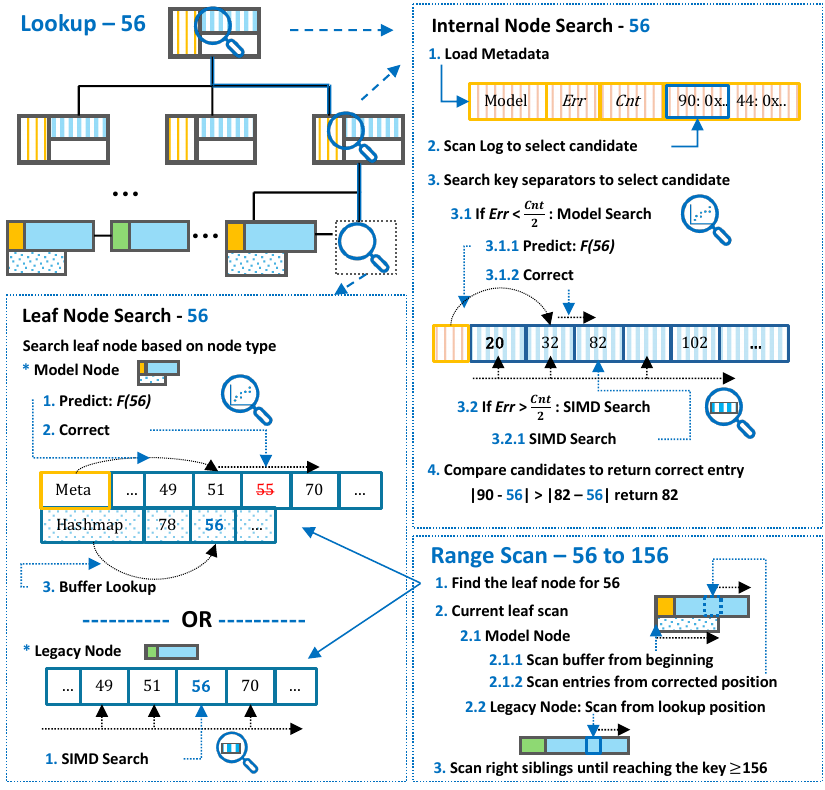} 
    \caption{Search of \opttree{}}
    \label{fig:search}
\end{figure}

Point lookups in \opttree{} are performed via a depth-first traversal from the root to a target leaf node, as illustrated in \cref{fig:search} with a query key $k_q = 56$. At each internal node, \opttree{} consults both the primary child list and a small log of newly added child node pointers to identify the lower-bound candidate node whose key range provides the tightest lower bound for the $k_q$. For our example search for key 56, the algorithm first performs a linear scan on the log. Given its small size (typically \(\leq10\%\) of the fanout \(f\)), the overhead of this scan is negligible. This scan finds a potential candidate node whose maximum key is 90. \rtwo{For the primary child list, the linear model's error is evaluated. If the error is less than half the child list's fanout, the model predicts \(k_q\)’s position, followed by a localized correction search around the predicted position. Otherwise, a SIMD-optimized linear search is used across the child list to ensure search efficiency.} In our example, this identifies the second candidate node whose maximum key is 82. Finally, \opttree{} compares the candidates. Between the candidate node ending at 90 (from the log) and the one ending at 82 (from the primary list), the latter provides a tighter lower bound for the key 56 and is selected for the next step in the traversal. 

Upon reaching a leaf node, the search method adapts to the node type. For a model-based leaf node, \opttree{} checks if $k_q = 56$ falls within the model’s key range. In the miss scenario depicted for key 56, this initial scan does not locate the key. The node's buffer is then queried using a hashmap, with \(O(1)\) complexity. If the node is undergoing concurrent retraining, an index-level buffer associated with the retraining process is also checked. For a legacy leaf node, a single, SIMD-accelerated linear search is performed across its sorted data array to locate the entry for key 56. Once a matching key-value pair \(\langle k_q, v_q \rangle\) is found, it is returned immediately; otherwise, an empty result is returned.

\subsubsection{Range Query} \label{sub_sec:range_query}

A range query extends the point query process to retrieve key-value pairs within a range \([k_l, k_u]\), such as $[56, 156]$ in \cref{fig:search}. The process begins with a point query to locate the leaf node containing the first key greater than or equal to \(k_l=56\), identifying its storage position \(p_{k_l}\). Subsequent steps depend on the leaf node type.

For a model-based leaf node, if \(k_u = 156\) is at most the maximum key in the node's data list (i.e., $k_u$ is equal or less than the last key in the node), \opttree{} collects entries from \(p_{k_l}\) to the position of the last key in range \([k_l, k_u]\). To ensure completeness, the node's buffer is also scanned for recently inserted key-value pairs within \([k_l, k_u]\). \rone{If the range covers data within the buffer, a local sorting step is performed. This operation is confined to the results retrieved from the current leaf, merging the sorted entries from the data list with the unsorted ones from the buffer to guarantee correctly ordered output.} If \(k_u\) exceeds the maximum key of the node (i.e., $k_u$ is greater than the last key in the node), all entries from \(p_{k_l}\) to the end of the data list are collected. \opttree{} then traverses to the next leaf node via its sibling pointer, repeating this process until \(k_u=156\) is covered or the end of the data is reached.

For a legacy leaf node, \opttree{} scans the sorted data list from \(p_{k_l}\), collecting entries until a key exceeds \(k_u=156\) or the list ends. Similar to the process in model-based nodes, if the range extends beyond the current node, the operation proceeds to the next leaf node via the sibling pointer until the query is no longer satisfied.

\subsection{Updates} \label{subsub_sec:updates}

Update operations in \opttree{}, encompassing insertions and deletions, are optimized for efficiency across node types. At internal nodes, \opttree{} employs a mechanism based on gap array and log to ensure efficient update performance. 
At the leaf layer, model-based leaf nodes employ an optimized buffer mechanism to defer the need for model retraining while keeping the search efficiency. Although the buffer stores data in a vector, our hash-based optimization enables $O(1)$ complexity for lookups, insertions, and deletions within it.

\begin{algorithm}[t]
\caption{\rv{\opttree{} Updates}}\label{alg:insertion}
\footnotesize
\SetKwFunction{FMain}{insert}
\Fn{\FMain{$k$, $v$}} {
	\KwIn{Insert key $k$ and value $v$}
	$node \leftarrow tree.root$\;
    \While{\FuncSty{isLeaf}(node) = false}{ \label{alg_line:traverse_start}
        \tcp{Check if the node is under retraining}
        \If{node.underRetrain = true} {
            $tree.\mathit{log}[node].\FuncSty{push}(ins, \langle k, v\rangle)$\; \label{alg_line:check_retrain_and_ins}
            \KwRet;
        }
        \Else{
            $node \leftarrow node.\FuncSty{nextChild}(k)$\;
        }
    } \label{alg_line:traverse_end}
    \If {node is model-based leaf} {
        $slot \leftarrow node.\FuncSty{predict}(k)$\; \label{alg_line:model_predict}
        \If{$node.data[slot]$ is deleted}  { \label{alg_line:check_gap}
            $node.data[slot] \leftarrow \langle k, v \rangle$\; \label{alg_line:insert_index_gap}
        }
        \Else{
            $node.\mathit{buffer}.\FuncSty{push}(k, v)$\; \label{alg_line:insert_buffer}
            $node.\mathit{buffer}.hashmap[k] = |node.\mathit{buffer}| - 1$\; \label{alg_line:record_insert_buffer_idx}
            $tree.cost.\FuncSty{update}(node) \gets$ \textbf{start thread do} \; \label{alg_line:check_retrain}
        }
        $leaf.cnt \gets leaf.cnt + 1$
    }
    \Else {
        $node.\FuncSty{LegacyInsert}(k, v)$\; \label{alg_line:legacy_insert}
        $tree.cost.\FuncSty{update}(node)\gets$ \textbf{start thread do}\; \label{alg_line:check_optimize}
    }
}
\SetKwFunction{FMain}{delete}
\Fn{\FMain{$k$}} {
    \KwIn{Delete key $k$}
    Do point query to $k$ to find storage $leaf$ and position $pos$ \;
    Log deletion if find nodes in the path are under retraining\;
    \If{$leaf$ is model leaf} {
        \If{$pos$ is in $D$} {
            \tcp{Mask data}
            $\FuncSty{mask}(leaf.data[pos])$\;
            \label{alg_line:mask_data}
            $leaf.cnt \gets leaf.cnt - 1$
        }
        \Else {
            $k_{last} \gets leaf.\mathit{buffer}.last.key$\; \label{alg_line:buffer_del_start}
            \FuncSty{swap}($leaf.\mathit{buffer}[pos]$, $leaf.\mathit{buffer}.last)$\;
            $leaf.\mathit{buffer}.hashmap[k_{last}] = pos$\;
            $leaf.\mathit{buffer}.hashmap.\FuncSty{remove}(k)$\;
            $leaf.\mathit{buffer}.\FuncSty{resize}(|leaf.\mathit{buffer}| - 1)$\; \label{alg_line:buffer_del_end}
        }
        
    }
    \Else {
        $leaf.\FuncSty{LegacyDelete}(k)$\;
    }
    \If{$leaf$ is underflow} {
        Do balancing operation\;
    }
}
\end{algorithm}

\begin{algorithm}[t]
\footnotesize
\caption{\rv{Internal Node Operations}}\label{alg:internal_insertion}
\SetKwFunction{FMain}{internalNodeInsert}
\Fn{\FMain{$inner$, $node_p$}} {
	\KwIn{Internal node $inner$  and pushed up node $node_p$}
    $slot \gets inner.\FuncSty{predict}(node_p.key)$\; \label{alg_line:inner_ins_predict}
    \If{inner.K[slot] is empty or deleted} {
        $inner.K[slot] \gets node_p.key$\; \label{alg_line:inner_ins_gap_s}
        $inner.P[slot] \gets node_p.ptr$\; \label{alg_line:inner_ins_gap_e}
    } 
    \Else{
        $inner.log.\FuncSty{append}( \langle node_p.key, node_p.ptr \rangle)$\; \label{alg_line:inner_ins_log}
    }
    $inner.cnt \gets inner.cnt + 1$\;
    \If{inner.cnt $> f$} {
        split $inner$ to $\langle inner_l, inner_r\rangle$\; \label{alg_line:inner_ins_overflow_s}
        \FuncSty{internalNodeInsert}$(inner_l.parent, inner_r)$\;
        Retrain $inner_l$ and $inner_l$ models, and remap $\gets$ \textbf{start thread do}\; \label{alg_line:inner_ins_overflow_e}
    }
}
\end{algorithm}

\begin{figure*}[tp]
    \centering
    \includegraphics[width=1\linewidth]{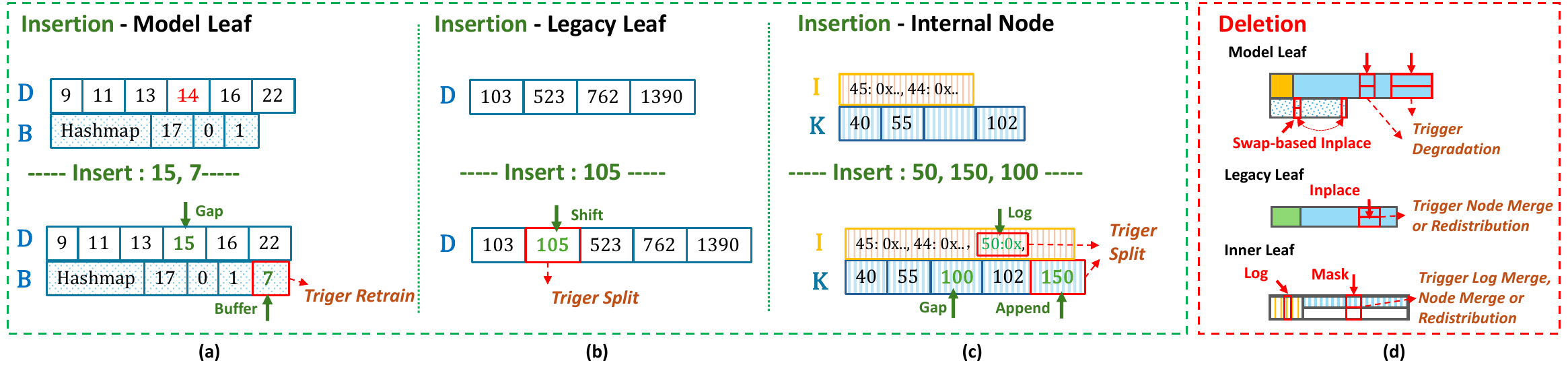} 
    \caption{Updates of \opttree{}}
    \label{fig:update}
\end{figure*}

\subsubsection{Insertion of New Data}

Algorithm~\ref{alg:insertion} describes the insertion process in \opttree{} for a new data point \(o = \langle k, v \rangle\). \opttree{} first locates the target leaf node by a point lookup, as detailed in Section~\ref{sub_sec:point_lookups}. During traversal, if a node is marked as undergoing retraining, \opttree{} places \(o\) in the index-level buffer and defers insertion until retraining is complete (Line \ref{alg_line:check_retrain_and_ins}). Upon reaching the leaf node, the insertion strategy depends on the node type.

For a model-based leaf node, \opttree{} first finds the insertion slot for \(o\) using model prediction and correction based on key \(k\) (Line \ref{alg_line:model_predict}). If the slot is marked as deleted (e.g., the slot for key 14 in \cref{fig:update}a), it can be reused without affecting the model's accuracy (Line \ref{alg_line:insert_index_gap}). Otherwise, \opttree{} appends \(o\) to the node’s buffer (e.g., key 7 in the example), placing it at the end and using a lightweight hashmap to track \(k\)'s position (Lines \ref{alg_line:insert_buffer}-\ref{alg_line:record_insert_buffer_idx}). This design can achieve \(O(1)\) complexity for buffer queries, insertions, and deletions. For a legacy leaf node, \opttree{} follows the standard B+-tree protocol (\cref{fig:update}b). The new object \(o\) is inserted directly into the sorted data list at the appropriate position. If the node overflows, it is split immediately. Finally, following any insertion, the \opttree{}'s cost model is updated (Lines \ref{alg_line:check_retrain} and \ref{alg_line:check_optimize}, to be detailed in \cref{sub_sec:retrain}). This allows \opttree{} to continuously evaluate whether a node requires recalibration, such as model retraining or legacy leaf node transformation.

\subsubsection{Deletion of Existing Data} \label{subsub_sec:deletion}
\cref{fig:update}d illustrates the deletion in \opttree{}. For model-based leaf nodes in \opttree{}, deletions of data integrated into the learned model use a mask-based strategy. \opttree{} locates the target entry via a point query, as described in Section~\ref{sub_sec:point_lookups}, and flags it as ``deleted'' by setting the flag bit of its key through a bitwise operation (Line \ref{alg_line:mask_data} in \cref{alg:insertion}). This preserves the key's position, maintaining the model's data distribution without requiring retraining. If the active data count falls below a predefined threshold $\alpha$, \opttree{} recalibrates by converting the model-based leaf node into one or more legacy leaf nodes to reduce index size, as will be detailed in Section~\ref{sub_sec:retrain}. Buffer deletions in model-based leaf nodes achieve \(O(1)\) complexity: \opttree{} swaps the target entry with the last active element, decrements the buffer size, and updates the hashmap to reflect the new position of the swapped element, avoiding a linear-time data movement operation (Lines \ref{alg_line:buffer_del_start}-\ref{alg_line:buffer_del_end}).

For legacy leaf nodes, \opttree{} performs in-place deletions using B+-tree-inspired procedures. If a deletion reduces the node's entries below the fanout \(f\), \opttree{} triggers a merge with an adjacent legacy leaf sibling or redistributes entries between them to maintain performance stability.


\subsubsection{Internal Node Updates} \label{sub_sec:internal_node_updates}

In \opttree{}, internal node updates occur when child nodes split (e.g., a leaf node splits or an internal node exceeds its fanout \(f\)) or merge (e.g., a node is merged into another). These processes are illustrated in Figure~\ref{fig:update}c and Algorithm~\ref{alg:internal_insertion}.

An insertion is triggered when a child node splits and "pushes up" a new key-pointer pair to an internal node. \opttree{} first uses the internal node's model to predict the target position for the new key (Line \ref{alg_line:inner_ins_predict}). If this position is a gap, the new entry is inserted directly (e.g., key 100 in \cref{fig:update}c). If the gap is occupied, the entry is instead appended to the log (e.g., key 50 in the example). This strategy avoids a costly $O(f)$ data movement. When the total number of child nodes in both the key-pointer (\(K\)-\(P\)) list and the log exceed \(f\), the internal node splits   into two B+-tree-inspired nodes. The newly created right node is then inserted into the parent node. Following the split, \opttree{} retrains the models for the new nodes in a background thread. Once training is complete, we scale the models' slopes and remap the child nodes, which can create new gaps for future insertions and enhances model precision (Lines \ref{alg_line:inner_ins_overflow_s}-\ref{alg_line:inner_ins_overflow_e}).

For deletions, triggered by child node merges, \opttree{} applies a mask-based deletion to mark the child as deleted in the \(K\)-\(P\) list, preserving gaps for future insertions, similar to model-based leaf nodes (Section~\ref{subsub_sec:deletion}). If the child is in the log, it is removed immediately. If the number of child nodes falls below \(f\), \opttree{} triggers a merge or redistribution with an adjacent sibling to maintain index balance and performance stability.

\begin{algorithm}[t]
\footnotesize
\caption{\rv{\opttree{} Model-based Leaf Node Retraining}}\label{alg:retrain}
\SetKwFunction{FMain}{retrainModelLeaf}
\Fn{\FMain{$node$}} {
	\KwIn{Model node $node$ needs to be retrained}
    $curr \gets node$\; \label{alg_line:snapshot_affected_nodes_start}
    \tcp{Maximum number of new nodes} 
    $\sigma \gets \lceil \frac{|\mathit{curr.data}| + |\mathit{curr.buffer}|}{f}\rceil - 1 $\; \label{alg_line:get_max_pushup} 
    \tcp{Snapshot nodes may affected by retraining}
    $\mathit{snapshotPath} , $ $\mathit{originalPath} \leftarrow []$; $\mathit{isFinished} \gets false$\; 
    \While{curr $\neq$ null \textbf{and} $\mathit{isFinished} = false$}{ 
        $\mathit{isFinished} \gets \sigma \leq 0$\;
        $curr.underRetrain \gets true$\; \label{alg_line:mark_under_retrain}
        $\mathit{snapshotPath}.\FuncSty{push}(\mathit{\FuncSty{copy}(curr)})$\;
        $\mathit{originalPath}.\FuncSty{push}(curr)$\;
        $curr \gets curr.parent$\;
        $\sigma \gets \lceil \frac{\sigma - (f - |\mathit{curr.childList}|)}{f} \rceil$\;
    } \label{alg_line:snapshot_affected_nodes_end}
    Create a log at index for updates for the subtree\;
    \tcp{Update pointer linkage of snapshot nodes}
    \FuncSty{UpdatePtrLink}($\mathit{snapshotPath}$)\; \label{alg_line:update_ptr}
    $\mathit{node} \gets \mathit{snapshotPath}.front$\;
    $\mathit{node}.\mathit{data} \gets$ \FuncSty{SortMerge}$(\mathit{node}.\mathit{data}, \mathit{node}.\mathit{buffer})$\; \label{alg_line:merge_buffer}
    $\mathit{node}.\mathit{model} \gets$ \FuncSty{newModel}$(\epsilon)$\; \label{alg_line:create_new_model}
    
    $cnt \gets 0$\;
    \tcp{Attempt to retrain the model}
    \For{$d \in node.data$}{ \label{alg_line:retrain_start}
        $\mathit{isAdded} \gets \mathit{node}.\mathit{model}.\FuncSty{addPoint}(d, cnt)$\;
        \If{$\mathit{isAdded}$ = false} {
            Split model leaf $node$ at $cnt$\ into $\langle node_l, node_r\rangle$\;
            \FuncSty{internalNodeInsert}($node_l.parent$, $node_r$\; \label{alg_line:split_node}
            \If{$cnt < \alpha$} {
                covert $node_l$ to legacy leaves\; \label{alg_line:convert_to_legacy}
            }
            $node \gets node_r$\;
            $cnt \gets 0$\;
        } 
        \Else{
            $cnt \gets cnt+1$\;
            \If{$cnt \geq \beta$} { 
                Split model leaf $node$ at $cnt$\ into $\langle node_l, node_r\rangle$\; \label{alg_line:split_at_beta}
                \FuncSty{internalNodeInsert}($node_l.parent$, $node_r$)\;
                $node \gets node_r$\;
                $cnt \gets 0$\;
            }
        }
    } \label{alg_line:retrain_end}
    Replace $\mathit{snapshotPath.last}$ with original index\; \label{alg_line:replace_subtree}
    
    \FuncSty{FreeNodes}($\mathit{originalPath}$)\; \label{alg_line:free_original_nodes}
    Execute all updates in $tree.\mathit{buffer}[originalPath.last]$\; \label{alg_line:insert_index_buffer_data}
}
\end{algorithm}

\subsection{Recalibration} \label{sub_sec:retrain}
\subsubsection{Cost Model}
\opttree{} incorporates a cost model to determine the optimal timing for retraining a model-based leaf node. Operating concurrently in the background, this model dynamically analyzes the frequency of operations  and the volume of buffered data associated with each model-based leaf node. This approach balances the performance gains from a more accurate model against the computational overhead of retraining. The decision to retrain is governed by two distinct triggers: a query-driven active trigger and a buffer-driven passive trigger.

\noindent\textit{Query-Driven Retraining (Active Trigger).}
This active trigger starts retraining when a model-based leaf node is both frequently queried and has accumulated a sufficient volume of new data. When a leaf node becomes a query ``hotspot'' and its buffer contains enough new records, the existing model may not accurately represent the data distribution, making it a prime candidate for an update.

A leaf node $l$ is flagged for retraining when its query count, $Q_l$, within a recent time window $T_q$ and its current buffer size, $B_l$, both exceed their respective thresholds, $Q_{th}$ and $B_{th}$:
    $Q_l \ge Q_{th}$, 
    $B_l \;\ge\; B_{th}$. 
%
%
When this condition is met, \opttree{} proactively initiates the retraining process. It merges the buffered entries into the data list of the leaf node and refits the model on the updated, unified data. This ensures that frequently accessed leaf nodes maintain high accuracy and search efficiency.

\noindent\textit{Parameter Tuning for Active Trigger}. 
The effectiveness of the active trigger hinges on the appropriate tuning of its parameters: the query frequency threshold $Q_{th}$, the buffer size threshold $B_{th}$, and the time window $T_q$. These parameters should be set to optimize the trade-off between the performance degradation duo to outdated models and the computational cost of retraining.

The decision to retrain should be made when the expected future performance gain outweighs the immediate, one-time cost of the retraining operation. We can formalize this relationship as follows. Let $C_{retrain}$ represent the fixed, measurable cost of retraining a leaf node, as the volume of update data is usually much less than the original data. This cost includes merging the buffer and retraining the model.
The performance benefit of retraining comes from moving $B_l$ entries from the inefficient, unordered buffer into the efficient, model accelerated data list. Let $c_{\mathit{buffer}}(B_l)$ denote the average cost of searching for a key within a buffer of size $B_l$, and let $c_{model}$ represent the cost of a search in the  data list using the model. The performance gain per query after retraining is given by the difference between these costs:
\begin{equation}
\Delta c = c_{\mathit{buffer}}(B_l) - c_{model} \nonumber
\end{equation}

The cost to scan the buffer is roughly proportional to its size, i.e., $c_{\mathit{buffer}}(B_l) \propto B_l$. In contrast, $c_{model}$ is primarily related to the model error bound $\epsilon$, which is very low and can be treated as a constant.
To estimate the total future benefit, we use the recently observed query count $Q_l$ over the time window $T_q$ as a predictor for future query arrivals. The total expected search cost savings, $C_{gain}$, before model retraining can be approximated by:
\begin{equation}
    C_{gain}\approx Q_l\cdot \Delta c = Q_l \cdot (c_{\mathit{buffer}}(B_l) - c_{model}) \nonumber
\end{equation}

Thus, retraining should be triggered  if and only if the expected gain surpasses the cost:
\begin{equation}
    C_{gain} > C_{retrain} \nonumber
\end{equation}

Substituting in the terms gives the core decision boundary:
\begin{equation}
    Q_l \cdot (c_{\mathit{buffer}}(B_{th}) - c_{model}) > C_{retrain} \nonumber
\end{equation}

Consequently, the cost model can adaptively tune the parameters $Q_{th}$ and $B_{th}$ by monitoring retraining and buffer scan costs, thereby maintaining the effectiveness of the active trigger.

For the time window size $T_q$, which is a predefined parameter by the user. It controls the stability and responsiveness of the query frequency measurement. A small $T_q$ allows the system to react quickly to workload spikes but may be sensitive to noise. Conversely, a large $T_q$ provides a more stable, long-term average but is slower to adapt to changes in data access patterns. The optimal value for $T_q$ depends on the expected volatility of the workload.

\noindent\textit{Buffer Overflow Retraining (Passive Trigger).}
Passive trigger addresses the scenario of buffer overflow. When the buffer of a model-based leaf node reaches its maximum capacity $\tau$, retraining becomes mandatory to integrate the buffered entries into the main data structure. Formally, retraining is initiated for a leaf node $l$ if its buffer size $B_l$ reaches this capacity, regardless of its query frequency:
\begin{equation}
    B_l \;\ge\; \tau \nonumber
\end{equation}
In this scenario, the data from the buffer is merged into the data list, and the model is subsequently refitted on the merged data. This trigger serves as a safeguard, preventing unbounded buffer growth that would otherwise degrade search performance by necessitating linear scans over a large, unsorted set of buffered records.

\begin{figure}[t]
    \centering
    \includegraphics[width=0.65\linewidth]{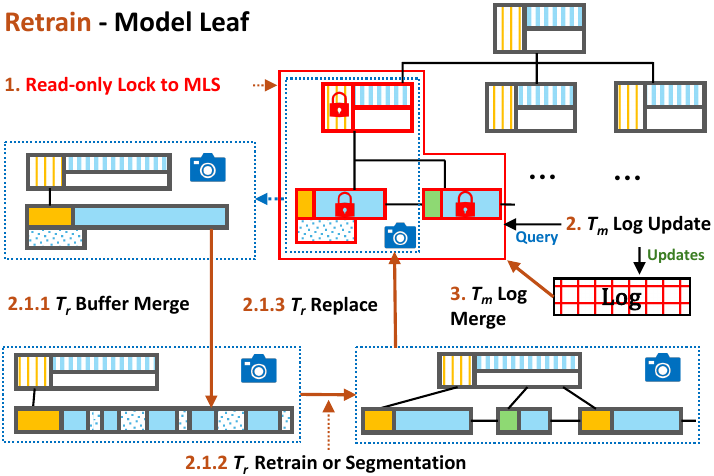} 
    \caption{Retraining of \opttree{}}
    \label{fig:retrain}
\end{figure}

\subsubsection{Model-based Leaf Retraining} \label{sub_sub_sec:model_leaf_retrain}
The balanced tree structure of \opttree{} facilitates index modifications, such as the replacement of specific subtrees. This inherent structural property is crucial for minimizing interference with concurrent operations accessing other regions of the index. To avoid blocking operations related to subtree reconstruction, we leverage modern multi-threaded hardware to design a non-blocking retraining mechanism based on Read-Copy-Update (RCU)~\cite{mckenney2001read}. This approach allows nodes to be updated in a copy and atomically replaced without impeding queries.
\iftechreport
\rthree{All operations of \opttree{} are executed under the Linux kernel RCU mechanism, using \textsf{rcu\_read\_lock()}, and \textsf{rcu\_read\_unlock()}, and \textsf{rcu\_dereference()} to safely access pointers. Once the modifications are complete, a pointer to the old version is safely and atomically swapped to point to the new version using \textsf{synchronize\_rcu()} and \textsf{rcu\_assign\_pointer()}.}
\else
Specifically, updates are prepared on a copy of the relevant portion of the index.\rthree{\footnote{\rthree{More details could be found in \cite{fullversion}}}}
\fi
This approach ensures that concurrent read operations can traverse the index without acquiring locks and without observing inconsistent, intermediate states of the data structure. Consequently, \opttree{} can update leaf nodes and their corresponding paths to internal nodes with minimal disruption to ongoing operations, thereby maintaining high availability and throughput.


As model-based leaf nodes have no gaps in data storage unless deletions occur, new insertions must be temporarily recorded in the buffer associated with the leaf node. To maintain search efficiency, a periodic retraining process integrates these buffered data with the existing leaf data. The procedure for the retraining is detailed in \cref{alg:retrain} and illustrated in \cref{fig:retrain}.

The retraining process for a target model-based leaf node begins in a separate thread $T_r$, which creates a snapshot of the potential affected path (PAP) from the leaf node up to the highest node that might be altered by child node splits (Lines \ref{alg_line:snapshot_affected_nodes_start}-\ref{alg_line:snapshot_affected_nodes_end}). During this snapshot, \opttree{} employs RCU to ensure safe traversal and snapshotting of the live index structure, even amidst concurrent modifications by other threads. This RCU-protected bottom-up snapshotting process considers the worst-case scenario where the model-based leaf node might split entirely into legacy leaf nodes. \opttree{} computes the maximum potential nodes that could be pushed up (Line \ref{alg_line:get_max_pushup}) and identifies the full PAP whose nodes will be copied for the snapshot.
Once the PAP is snapshotted, this copied path forms the basis for constructing the new subtree version and will be operated on by the retraining thread $T_r$. The original nodes in the main index corresponding to this PAP remain accessible to other concurrent operations, consistent with RCU. We refer to the subtree rooted at the highest node in the PAP snapshot as the minimal locked subtree (MLS). To capture concurrent modifications to the MLS while $T_r$ is working, incoming updates to this specific subtree are temporarily recorded in a dedicated log. 

Within $T_r$, the retraining process begins by merging the buffered data of the leaf node with its existing data entries (Line \ref{alg_line:merge_buffer}).  A streaming linear fitting process is then applied to this merged dataset, ordered by keys, subject to a predefined fitting error threshold $\epsilon$ (Lines \ref{alg_line:retrain_start}-\ref{alg_line:retrain_end}). If a data entry cannot be fitted within $\epsilon$, the current model segment ends, a new leaf node is formed with the successfully fitted data, and this new node is incorporated into the PAP snapshot (Line \ref{alg_line:split_node}). If a segment of fitted data does not meet a minimum entry count $\alpha$ required for a model-based leaf node, it is converted into one or more legacy leaf nodes within the snapshot (Line \ref{alg_line:convert_to_legacy}). Furthermore, even if the data fits the model well, a model-based leaf node is split if its entry count exceeds a maximum capacity $\beta$  (Line \ref{alg_line:split_at_beta}). This capacity constraint prevents excessively large nodes, which could lead to increased retraining overhead or significant structural changes if data distributions shift dramatically.


Upon processing all merged data, the modified PAP snapshot contains the newly retrained version of the subtree. \opttree{} then atomically installs this new subtree into the main index by updating the parent pointer to the root of the MLS, using an RCU-safe pointer swap (Line \ref{alg_line:replace_subtree}). Following this, the nodes from the previous version of the subtree are freed after a grace period, ensuring no active readers are still using them (Line \ref{alg_line:free_original_nodes}). Finally, the updates captured in the log of the MLS during the retraining period are applied to the newly installed subtree (Line \ref{alg_line:insert_index_buffer_data}). This two-phase approach ensures data consistency and high index availability throughout the retraining process.

\subsubsection{Legacy Leaf Node Transformation}

Legacy leaf nodes in \opttree{} can be transformed into model-based leaf nodes via two distinct mechanisms: forward merging and backward merging. Forward merging involves incrementally training an adjacent existing model-based leaf node with the data from a legacy leaf node, while backward merging consolidates multiple consecutive legacy leaf nodes into a new, single model-based leaf node. Concurrently, the cost model within \opttree{} continuously monitors legacy leaf nodes, computing their linear regression coefficients upon each update and tracking the position and aggregate length of consecutive legacy leaf node sequences in the index structure.


Forward merging is initiated when the linear regression coefficients of a legacy leaf node closely match those of its preceding model-based leaf node. In such cases, an attempt is made to incrementally train the existing model-based leaf node. Similar to the standard retraining procedure for model-based leaf nodes, a snapshot of potentially affected subtree nodes is taken. Using this snapshot, incremental learning is attempted: data from the legacy leaf node is streamed to the model-based leaf node for training. If the model-based leaf node can accommodate both its original data and the data from the legacy leaf node while maintaining its error within $\epsilon$, the legacy leaf node is merged into the model-based leaf node, and the main index is updated accordingly. To ensure safe concurrent access during this merging, the update to the main index is managed using RCU. If the error constraint is violated, the merging attempt is aborted.


Backward merging is considered when the cost model identifies a sequence of multiple consecutive legacy leaf nodes that (a) exhibit similar linear regression coefficients and (b) whose combined data volume meets the minimum requirement for creating a model-based leaf node. The merging process operates on snapshots of these legacy leaf nodes. A piecewise linear approximation (PLA) model is then employed to fit the  data from these legacy leaf nodes, with the fitting error constrained to be within $\epsilon$. If the data can be collectively fitted by the model under this error threshold, these legacy leaf nodes are converted into a new model-based leaf node, and the main index is subsequently updated to reflect this change. This structural update also leverages RCU, allowing concurrent read operations to proceed safely and without traditional locks while the index structure is modified.

\subsection{Bulk Loading} \label{sub_sec:bulk_load}

The bulk-loading algorithm in \opttree{} is designed to optimize the index structure by strategically partitioning data into leaf nodes. The goal is to ensure that the models within the upper-level internal nodes can achieve low fitting errors, thereby improving search performance without increasing the overall index height. However, given a dataset of size $N$, finding a globally optimal partition of the data (i.e., minimizing the average model error along every path) requires a dynamic programming approach with $O(N^3)$ complexity, which is hard to apply to large datasets. To address this challenge, we propose an approximate algorithm that performs inter-level optimization during bulk-loading with $O(N)$ complexity.

Our bulk-loading process employs a bottom-up strategy, controlled by a key hyperparameter: the error tolerance $\delta$. This parameter enables localized optimization during data partitioning, allowing for more efficient organization of the index structure.
Given an input dataset of $N$ ordered keys, $K=\{k_i,k_2,\cdots,k_N\}$, the construction process scans this sorted sequence to build the index level-by-level, beginning at the leaf layer. For each parent node of the leaf layer, we first populate it with an initial set of child nodes. Specifically, we generate the first $\frac{f}{4}$ leaf nodes according to the data partitioning rules of leaf retraining described in \cref{sub_sec:retrain}. The partitioning keys of these leaf nodes are then used to fit an initial linear regression model, denoted as $\mathcal{F}$.
For each subsequent leaf node that can be formed under this parent, we introduce a $\delta$-bounded optimization. Instead of using a fixed approach, we define a candidate window of $\delta$ keys around the original partitioning key. The optimal partitioning key is chosen from this window to minimize its deviation from the parent's regression model $\mathcal{F}$. The deviation for a candidate key $k$ with an actual rank of $\mathcal{R}(k)$ is defined as a distance $\mathcal{D}(k) = \left| \mathcal{F}(k) - \mathcal{R}(k) \right|$. The key that minimizes this distance $\mathcal{F}(k)$ is selected as the final partitioning key and appended to the parent node. The model $\mathcal{F}$ is next updated in an online fashion using Recursive Least Squares (RLS) \cite{hayes1996statistical}.

This strategy is applied recursively to higher levels of the tree. When constructing an internal node at layer $i+1$, its children are the nodes at layer $i$.  The ``keys" used to build the model for the layer $i+1$ node are the partitioning keys chosen for the layer $i$ nodes. When an internal node at layer $i$ becomes full and requires splitting, we apply a similar tolerance-based approach. 
The process terminates once all keys from $K$ have been indexed, resulting in a complete and optimized \opttree{} index.

\noindent
\textit{Complexity Analysis.} Assuming the input dataset of $N$ keys is presorted, the construction of the leaf layer requires a single linear scan over the data, which is an $O(N)$ operation. Subsequently, building the linear model for each internal node involves scanning the partitioning keys of its child nodes. Since each key is used as a partitioning key for at most one parent node in the level immediately above it, the cumulative work for building all models across all layers is also proportional to $N$. Therefore, the total time complexity of \opttree{} bulk-loading algorithm is $O(N)$.

\rone{
\subsection{Discussion} \label{sub_sec:discussion}
\noindent
\textbf{Duplicate Keys.} \opttree{} could handle duplicate keys using a value list approach, a strategy common in traditional index implementations. The payload associated with duplicated key is a pointer to the head of a linked list, where each node in the list holds one of the values corresponding to the duplicate key. 

\noindent
\textbf{Multi-dimensional Keys.} While \opttree{} is designed as an ordered index for single-attribute numeric keys, it can support multi-attribute keys using a standard lexicographical mapping, similar to that of the B+-tree~\cite{comer1979ubiquitous}. Specifically, a composite key $\textbf{k} = (k_1,\cdots, k_m)$ is transformed into a single 1D key using an order-preserving function $P(\textbf{k})$. \opttree{} can then be built on these mapped keys. This approach is highly efficient for queries on the leading attribute. For example, a range predicate on the first column $k_1 \in (L, U)$ can be converted into a query for a single, contiguous interval $[P(L, min_{k_2,\cdots, k_m}), P(U, max_{k_2,\cdots, k_m})]$ on the mapped keys. \opttree{} then completes the search with a single range scan.
}

\begin{figure*}[tp]
    \centering
    \includegraphics[width=\linewidth]{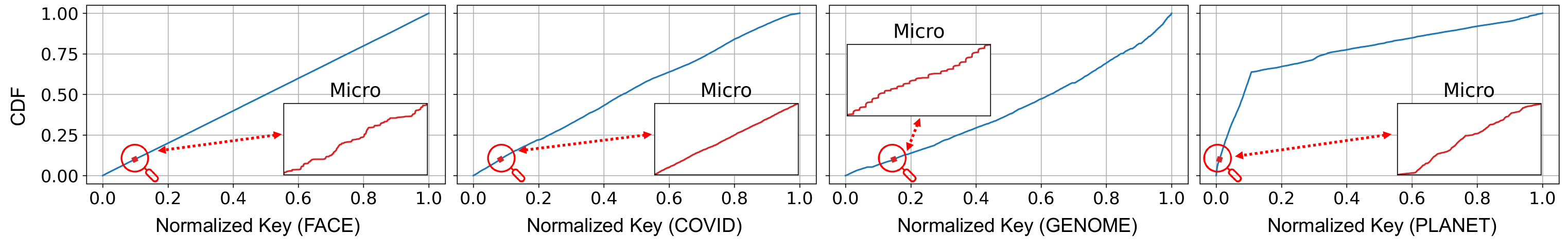} 
    \caption{CDFs of \textsf{FACE}, \textsf{COVID}, \textsf{GENOME} and \textsf{PLANET} Datasets}
    \label{fig:cdf_covid_genome}
\end{figure*}

\section{Evaluation} \label{sec:experiments}

\subsection{Experimental Setup} \label{sub_sec:exp_setup}
\noindent
\textbf{Datasets and Workloads.} We conduct experiments on six real-world datasets from SOSD~\cite{kipf2019sosd} and GRE~\cite{wongkham2022updatable}. All datasets consist of 200 million 64-bit unsigned integer keys, each paired with a random 64-bit unsigned integer value for experimental purposes. The following three datasets are selected from SOSD: \textsf{OSM} (uniformly sampled from OpenStreetMap locations), \textsf{FACE} (upsampled from Facebook user ID data), and \textsf{AMZN} (Amazon book sales popularity data). For GRE, we select the datasets \textsf{COVID} (uniformly sampled Twitter IDs labeled with COVID-19), \textsf{GENOME} (human chromosome gene location pairs), and \textsf{PLANET} (planet ID data derived from OpenStreetMap). The CDF of keys for the \textsf{FACE}, \textsf{COVID}, \textsf{GENOME}, and \textsf{PLANET} datasets is shown in \cref{fig:cdf_covid_genome}. 

For all datasets, unless otherwise specified, 20\% of the data is bulk loaded to construct the indexes. To simulate potential distribution shifts, a default balanced workload is applied, consisting of a sequence of random operations totaling 75\% of the dataset size. These operations include queries, insertions, and deletions in a 1:1:1 ratio. \rthree{For insertions, keys are drawn uniformly at random from the rest of the dataset, while deletions and queries are drawn uniformly from the dynamic set of keys currently stored in the index.} All queries are executed as range queries with a default match rate (\# data points falling within the range) of 256 results.

\noindent
\textbf{Metrics.} In our experiments, 
we evaluate two key performance metrics: the \textit{latency} of each operation (measured from submission to completion) and the \textit{throughput} of the index, expressed as the number of operations processed per second (ops). 
For index size, we record the \textit{memory usage} at fixed intervals throughout the workload execution and use the average of these measurements as the representative index size.

\noindent
\textbf{Indexes for Evaluation.} We compare our proposed \opttree{} with three general-purpose learned indexes previously discussed in \cref{sec:background_motivation}: ALEX,\footnote{https://github.com/microsoft/ALEX} PGM,\footnote{https://github.com/gvinciguerra/PGM-index} and LIPP.\footnote{https://github.com/curtis-sun/TLI/tree/main/competitors/lipp} 
Note that we exclude SWIX~\cite{liang2024swix} from our experiments because it is specifically designed for streaming data workloads and does not support the full set of operations required by general-purpose indexes. We also employ B+-tree\footnote{https://github.com/tlx/tlx/tree/master/tlx/container} as a baseline representing traditional indexing structures. To ensure a fair comparison, SIMD acceleration is enabled for all baselines. 

\iftechreport
\begin{figure}[t]
    \centering
    \includegraphics[width=.95\linewidth]{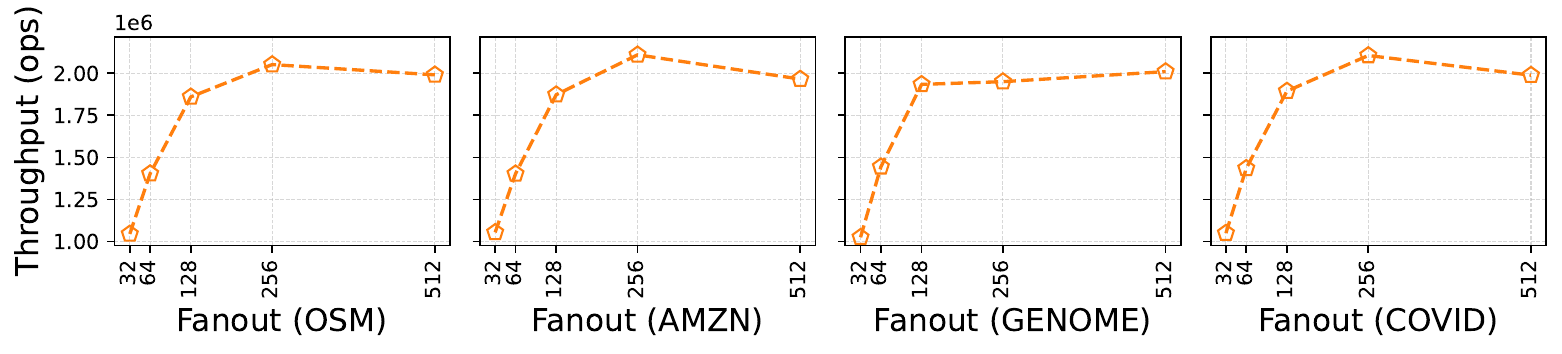} 
    \caption{Throughput of B+-tree on Different Fanout}
    \label{fig:exp_fanout}
\end{figure}
\fi

\rtwo{
\noindent
\textbf{Parameter Selection.}\iftechreport \ \else\footnote{\rtwo{Due to space constraints, a detailed explanation of the parameters and their selection is provided in our technical report~\cite{fullversion}.}}~\fi
\iftechreport
For B+-tree, we use a fanout of 256, which achieves the best performance across most datasets as shown in \cref{fig:exp_fanout}. For PGM, we sweep its error parameter from 4 to 4,096 and report the configuration yielding the lowest latency in each experiment. For \opttree{}, we set its parameters as follow:

\begin{itemize}

\item \textbf{Internal node fanout ($f=256$).} We set the fanout of \opttree{}’s internal nodes to 256 to enable a fair comparison with B+-tree. This choice also strikes a balance between node capacity, update overhead, and cache efficiency.

\item \textbf{Model-based leaf node minimum size ($\alpha = 512$) and maximum size ($\beta = 32,768$).} The parameters $\alpha$ and $\beta$ are derived from $f$ to control the capacity of model-based leaf nodes. Specifically, setting $\alpha = 2f$ ensures that \opttree{} does not create excessively small model-based leaf nodes, which would otherwise result in an unnecessarily large number of leaf nodes. Meanwhile, $\beta = f \times \frac{f}{2}$ guarantees that, even in the extreme case where a model-based leaf node completely splits into legacy leaf nodes, its parent will undergo at most one split. This design bounds the structural changes during recalibration and prevents drastic modifications to the index.

\item \textbf{Error bound of model-based leaf node ($\epsilon = 64$).} We set the maximum allowable error of a model-based leaf node to 64 to balance accuracy and efficiency. This threshold ensures that the model maintains reasonable prediction accuracy, while enabling cache alignment during the correction process after prediction, thereby improving query efficiency. 

\item \textbf{Buffer size of model-based leaf node ($\tau = 256$) and error tolerance of bulk loading ($\delta = 8$).} We set the maximum buffer capacity to $f$ to ensure that when a range query result overlaps with the buffer, its efficiency remains comparable to that of a B+-tree leaf node. The error tolerance parameter is introduced primarily to improve the efficiency of bulk loading. A smaller $\delta$ reduces the overhead of evaluating the impact of different keys on the parent node models during bulk loading, thereby accelerating the initial construction process.
\end{itemize}

We note that these parameters could be further optimized using automatic tuning tools like OpenTuner~\cite{DBLP:conf/IEEEpact/AnselKVRBOA14}, DDPG~\cite{DBLP:journals/corr/LillicrapHPHETS15}, or LITune~\cite{wang2025new}. However, as automated parameter tuning is orthogonal to the core contributions of this work, we leave such exploration for future work.

\else
For B+-tree, we use a fanout of 256, which achieves the best performance across most datasets. For PGM, we sweep its error parameter from 4 to 4,096 and report the configuration yielding the lowest latency in each experiment. For \opttree{}, we set $f = 256$, $\alpha = 2f = 512$, $\beta = f \times \frac{f}{2} = 32,768$, $\epsilon = \frac{f}{4} = 64$, $\tau = f = 256$, and $\delta=8$ in all experiments.
\fi
}

\noindent
\textbf{Environment.} All experiments are conducted on a workstation equipped with an AMD Ryzen 9950X and 64 GB of RAM, running Arch Linux with kernel 6.6.75. All algorithms are compiled with GCC 14.2.1, using the optimization flags \textsf{-O3 -march=native -flto}.

\begin{figure*}
    \centering
    \begin{minipage}{1\linewidth}
        \centering
        \includegraphics[width=0.98\linewidth]{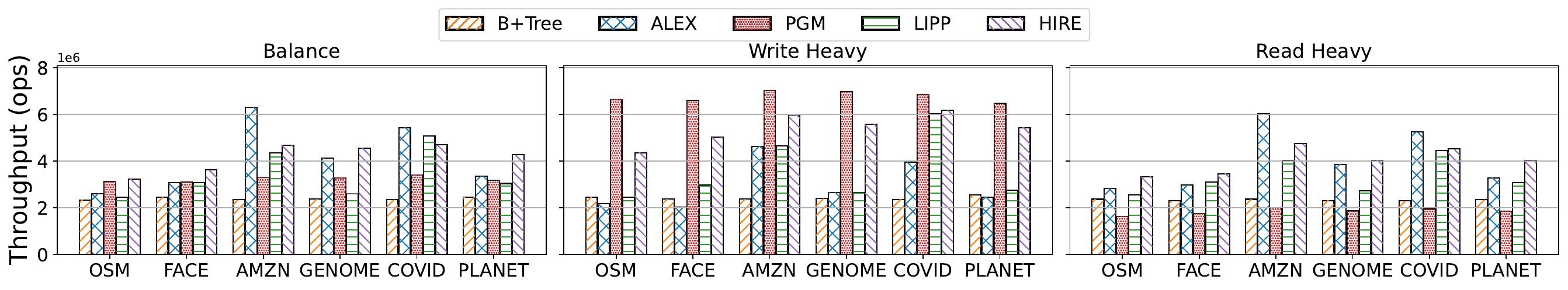}
        \caption{Throughput on Lookup Queries}
        \label{fig:exp_lookup}
    \end{minipage}
\end{figure*}

\subsection{Performance on Different Workloads}
This section compares the performance of various algorithms across multiple datasets and diverse workloads. In addition to the default balanced workload, two others are evaluated: write-heavy and read-heavy. The write-heavy workload emulates a write-intensive environment by adjusting the ratio of queries, insertions, and deletions to 1:8:1. Conversely, the read-heavy workload uses an 8:1:1 ratio to mimic a read-intensive environment. These workloads enable evaluation of performance variations stemming from the design decisions of each algorithm.

\subsubsection{Point Lookup Query}

We begin by comparing the throughput performance of each index on point lookup queries by setting the match rate to one result. Point lookup queries are fundamental to all other operations, and their performance critically impacts the overall efficiency of the index.

The results, shown in Figure \ref{fig:exp_lookup}, demonstrate that B+-tree maintains consistent performance across different datasets with distinct data distributions. In contrast, learned indexes can experience a throughput degradation exceeding $2.4\times$ when transitioning from model-friendly distributions (e.g., \textsf{AMZN}) to those more challenging to learn (e.g., \textsf{OSM}). Despite this, learned indexes generally outperform B+-tree on point lookup queries.

Contrary to the general perception of their update inefficiency, learned indexes exhibit robust performance under the write-heavy workload, primarily due to their buffering strategies. For example, PGM employs an LSM-tree as an index-level buffer, allowing direct insertion of new keys without searching the main structure, which achieves the highest throughput.
Conversely, finer-grained, data-level buffers, like that used by ALEX, require locating the correct position for each insertion, negatively impacting update throughput. This buffer granularity also introduces a trade-off between search and update performance.
Under the read-heavy workload,  this trade-off penalizes PGM, as its need to search multiple buffers results in the lowest throughput (only 84\% of B+-tree even on model-friendly distributions like \textsf{AMZN} and \textsf{COVID}). 
In contrast, ALEX's data-level buffer preserves data locality and confines searches to a single node, achieving significantly higher throughput -- $1.2\times$ and $2.5\times$ throughput of B+-tree on \textsf{OSM} and \textsf{AMZN}, respectively.

\opttree{} and LIPP demonstrate relatively stable performance across different workloads. LIPP leverages a precise, nearly error-free model for excellent search performance. To prevent model drift from insertions, LIPP splits data nodes to maintain accuracy, which however increases pointer traversal costs and causes update performance to vary with data distributions. 
By contrast, \opttree{} uses a hybrid buffering technique at different index levels, efficiently ingesting new keys while mitigating insertion costs. Its lightweight buffering mechanism enables low-cost searches, thereby yielding strong search performance. Even on \textsf{OSM}, the least linearly distributed dataset, \opttree{} excels under the read-heavy workload, achieving 1.2$\times$ throughput of ALEX, the best-performing baseline.


\begin{figure*}
    \centering
    \begin{minipage}{1\linewidth}
        \centering
        \includegraphics[width=0.46\linewidth]{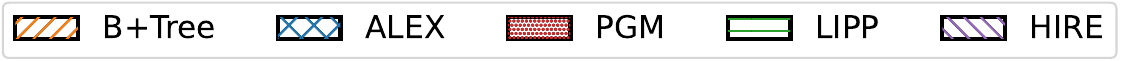}
    \end{minipage}
    \begin{minipage}{1\linewidth}
        \centering
        \includegraphics[width=\linewidth]{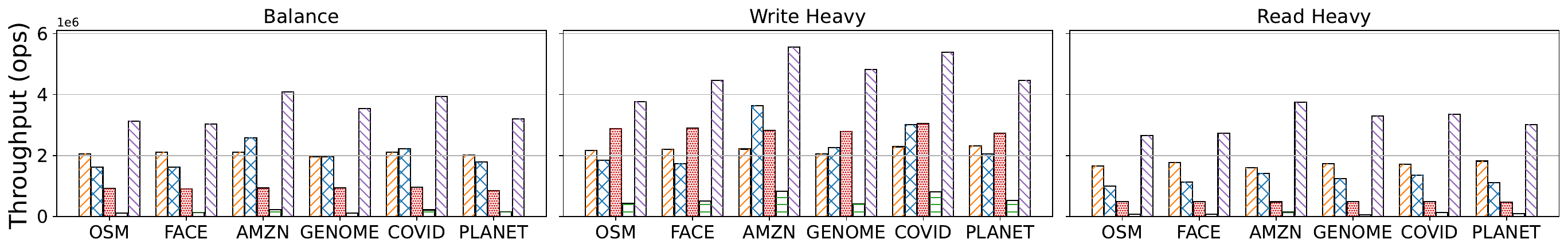}
        \caption{Throughput on Range Queries}
        \label{fig:exp_range}
    \end{minipage}
    \begin{minipage}{1\linewidth}
        \centering
        \includegraphics[width=\linewidth]{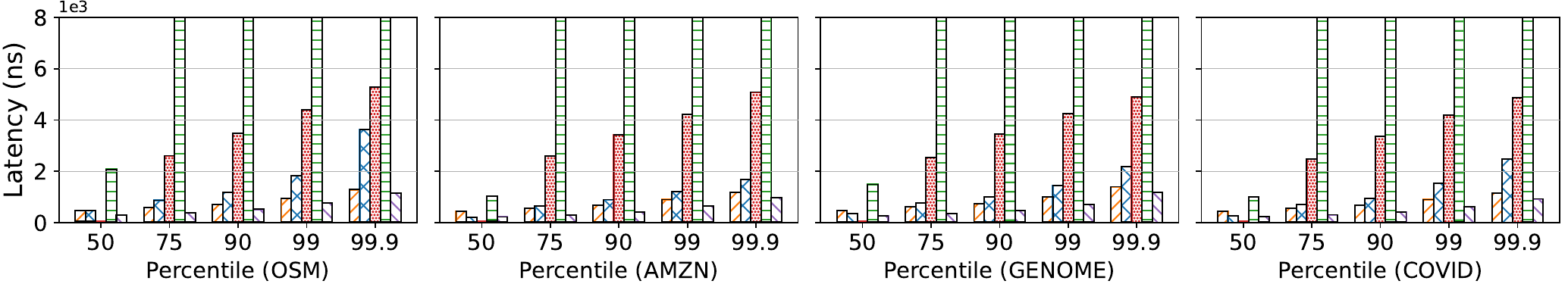}
        \caption{Tail Latency on Different Datasets}
        \label{fig:exp_tail}
    \end{minipage}
\end{figure*}

\subsubsection{Range Query}
In our experiments, range queries use the same settings as point lookup queries, except the match rate is fixed at 256 results. Evaluating learned indexes on range queries would reveal an additional trade-off factor: data locality. Data locality is crucial because, after the index locates the first key in a range, it must sequentially scan contiguous data until it encounters a key outside the specified range. 
Poor data locality can force the index to search multiple buffers or perform pointer jumping, significantly increasing query latency.

Figure \ref{fig:exp_range} shows the stable performance of B+-tree. For range queries, most state-of-the-art learned indexes fail to outperform B+-tree across different workloads and datasets. The performance degradation from point lookup to range queries in learned indexes arises from their inherent trade-off between data locality and update/lookup efficiency. This trade-off involves factors such as ALEX's sparse structure, PGM's index-level buffering, and LIPP's splitting strategy. Among these, LIPP exhibits the greatest performance decline, with a 24.4$\times$ performance degradation compared to point lookup queries, as its splitting strategy prevents data from being stored contiguously within a node, necessitating costly pointer traversals that reduce efficiency.
In contrast, B+-tree achieves robust performance as its buffer is located only at the end of each node, thereby preserving data locality and minimizing pointer traversals. 

\opttree{} consistently delivers the strongest range query performance. For example, it exceeds B+-tree by up to 1.9$\times$, $2.5\times$, and $2.4\times$ in throughput under the three different workloads, respectively. Compared to other learned indexes, \opttree{} outperforms LIPP by up to 41.7$\times$, ALEX by 2.7$\times$, and PGM by 7.9$\times$. 
This performance advantage stems from two key designs: cache-aligned lightweight buffers that reduce overhead and a compact data layout that preserves data locality within model-based nodes.
For less model‐friendly distributions like \textsf{OSM}, \opttree{} leverages legacy nodes to maintain data locality and ensure consistently high range query performance.



\subsection{Performance on Tail Latency}

Learned indexes often improve average performance by delaying expensive operations. However, tail latency is also critical as the deferred operations can negatively impact worst-case scenarios, 
leading to higher overall processing delays and greater uncertainty in index performance.

Figure \ref{fig:exp_tail} shows the 50th, 75th, 90th, 99th, and 99.9th percentile tail latencies for various indexes on four datasets. As percentiles increase, the tail latency of B+-tree rises gradually but remains relatively stable across datasets. In contrast, learned index baselines exhibit a more pronounced increase. Notably, LIPP has the highest tail latency for all datasets. PGM’s tail latency pattern remains consistent across datasets, 
likely due to its distribution-agnostic insertion and merge strategy. ALEX achieves relatively low tail latency in model-friendly distributions, but performs poorly in model-unfriendly scenarios, where degraded structural balance increases recalibration overhead. \opttree{} consistently exhibits low tail latency across all datasets. 
Its balanced design, non-blocking cost-driven recalibration mechanism, and proactive use of legacy leaf nodes for challenging distributions effectively minimize update costs.
In particular, on the challenging \textsf{OSM} dataset, \opttree{} reduces tail latency by up to 37\% compared to B+-tree, 68\% to ALEX, 85\%  to PGM, and 97\% to LIPP. On the model-friendly \textsf{AMZN} dataset, \opttree{} achieves tail latency reductions of up to 51\% compared to B+-tree, 56\% to ALEX, 88\% to PGM, and 98\% to LIPP.



\begin{figure*}
    \centering
    \begin{minipage}{0.6\linewidth}
        \centering
        \includegraphics[width=0.91\linewidth]{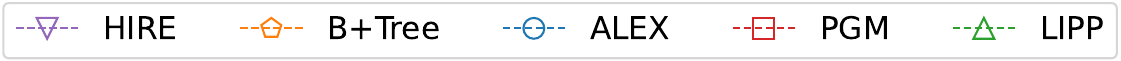}
    \end{minipage}
    \begin{minipage}{1\linewidth}
        \centering
        \includegraphics[width=0.98\linewidth]{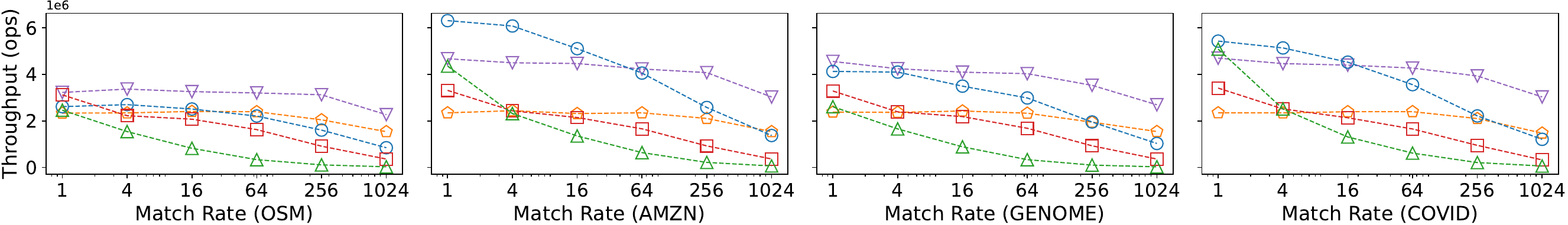}
        \caption{Throughput on Different Match Rates}
        \label{fig:exp_matching_rate}
    \end{minipage}
\end{figure*}

\subsection{Performance on Different Match Rates}
Range query performance is heavily influenced by the match rate. A lower match rate restricts scans to a single node, whereas a higher match rate requires traversing multiple nodes, with expensive pointer traversals. We vary the match rate from 1 to 1,024 and evaluate the indexes on different datasets, as shown in Figure \ref{fig:exp_matching_rate}. 

Generally, increasing the match rate decreases throughput due to larger scan ranges and increased random accesses when data locality is compromised. For example, ALEX underperforms on the \textsf{OSM} dataset because it creates more gaps within nodes and requires extra nodes to maintain its error threshold, increasing scan costs. Similarly, ALEX excels on the \textsf{AMZN} dataset at low match rates but degrades significantly at higher match rates due to bypassing additional gaps. LIPP achieves optimal performance at a match rate of 1 (i.e., point lookup) across all datasets. However, even a slight increase in the match rate substantially reduces LIPP's performance due to random accesses. In contrast, \opttree{} consistently demonstrates robust performance due to its compact leaf node structure with model-legacy duality, which effectively maintains data locality across different data distributions. On the challenging \textsf{OSM} dataset, \opttree{} achieves up to $1.5\times$ higher throughput than B+-tree and $2.7\times$ higher throughput than the best-performing learned index baseline, ALEX, at a match rate of 1,024. 

\begin{figure*}
    \centering
    \begin{minipage}{0.6\linewidth}
        \centering
        \includegraphics[width=0.91\linewidth]{figs/exp_varying_range_legend.pdf}
    \end{minipage}
    \begin{minipage}{1\linewidth}
        \centering
        \includegraphics[width=0.98\linewidth]{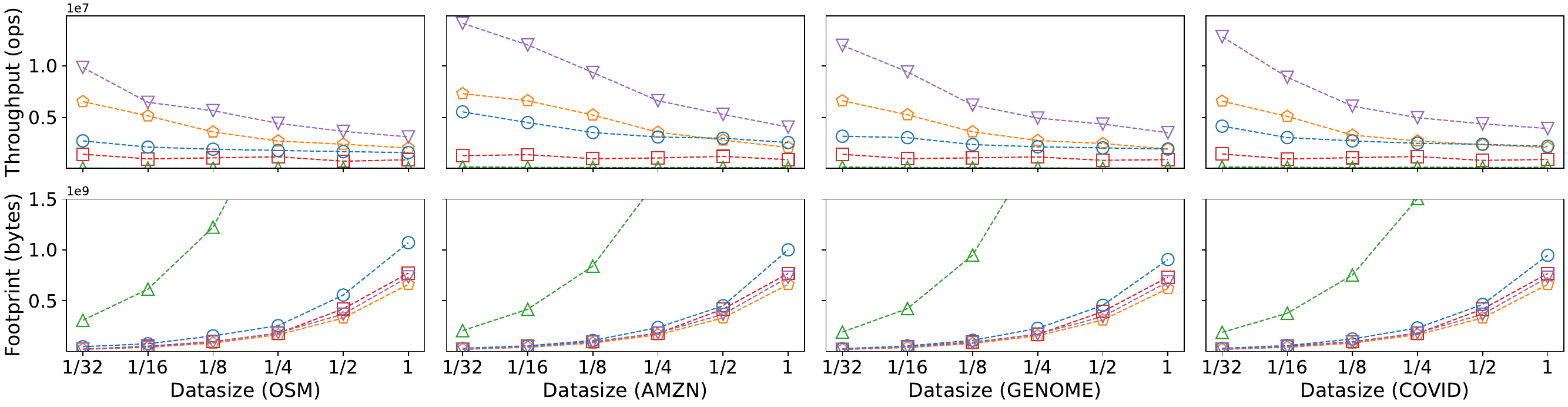}
        \caption{Throughput and Index Size on Different Datasizes}
        \label{fig:exp_varing_size}
    \end{minipage}
    \begin{minipage}{\linewidth}
        \centering
        \includegraphics[width=0.98\linewidth]{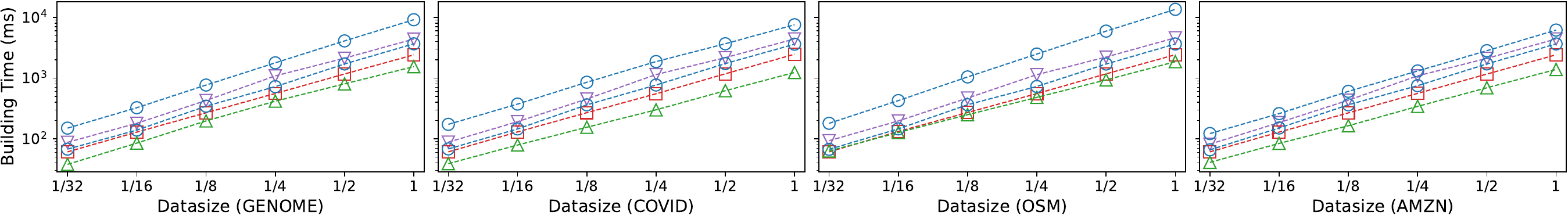} 
        \caption{\rone{Index Build Time on Different Datasizes}}
        \label{fig:building}
    \end{minipage}
\end{figure*}

\subsection{Performance on Scalability}
Figure \ref{fig:exp_varing_size} shows the throughput (top) and memory usage (bottom) of various indexes as dataset size increases. When the dataset size grows, all indexes experience performance degradation, with throughput decreasing and memory usage increasing. However, HIRE consistently outperforms all baselines in terms of throughput while maintaining memory consumption comparable to other learned indexes, demonstrating its superior scalability. Compared to the best baseline, B+-tree, \opttree{}'s index size is $1.1\times$ larger, but its throughput is $1.5\times$ higher.

\rone{
\subsection{Performance on Index Building} \label{sub_sec:exp_building}
For in-memory indexes, construction time is a critical performance indicator. We present the initial build times for various indexes across different dataset sizes in \cref{fig:building}. As observed, the build time of all algorithms grows roughly linearly with dataset size.

Among them, LIPP achieves the fastest construction time, outperforming the standard B+-tree by at least $1.8\times$. This is because LIPP's bulk loading process computes a minimal collision mapping (i.e., multiple keys are mapped to the same slot by the model) over the entire input, which allows it to generate long, contiguous arrays and thereby accelerates construction. This design, introduces a trade-off. While it accelerating the initial construction, these large contiguous segments can incur high split cost during future updates, particularly under data distribution shifts. PGM and B+-tree exhibit similar build times, as both rely on segmenting the data before constructing their upper-level index. The construction of \opttree{} is moderately slower, incurring a $1.2$-$1.5\times$ overhead compared to PGM. This is an expected trade-off for its strategy, which finds a more optimized index structure by exploring inter-level segmentation. ALEX also optimizes the structure during construction at the cost of build time. It consistently incurs the highest build time, taking up to $3.7\times$ longer than B + tree in our experiments. This significant overhead stems from its complex fanout tree design, which is particularly costly for non-linear datasets.
}

\subsection{Ablation Studies of \opttree{}}
\subsubsection{Effectiveness of Hybrid Nodes}
To validate the effectiveness of our proposed hybrid node design, we conduct an ablation study comparing the full \opttree{} index against a variant without legacy leaf nodes. We select the model-unfriendly \textsf{OSM} dataset and the model-friendly \textsf{AMZN} dataset to evaluate the performance impact across different data distributions. The results are shown in \cref{fig:exp_hybrid_nodes}. On \textsf{OSM}, the removal of legacy leaf nodes causes a significant 26\% drop in throughput, clearly demonstrating the importance of the hybrid leaf design for ensuring robustness on skewed data distributions. The hybrid node design allows \opttree{} to consolidate numerous small, hard-to-learn data segments into fewer, larger legacy leaf nodes,  significantly reducing the total number of leaf nodes in the index and thereby enhancing overall performance and throughput. On \textsf{AMZN}, the variant without legacy leaf nodes exhibits a $8\%$ decrease in throughput comparable to the full \opttree{}. For such model-friendly data, most segments are linear enough to form model-based leaf nodes naturally. 
Nevertheless, this variant of \opttree{} still benefits from the remaining components of the design,  making it to outperform other baselines.

\subsubsection{Effectiveness of Non-Blocking Recalibration}
\begin{figure}[t]
\begin{minipage}{0.8\linewidth}
    \centering
    \includegraphics[width=0.75\columnwidth]{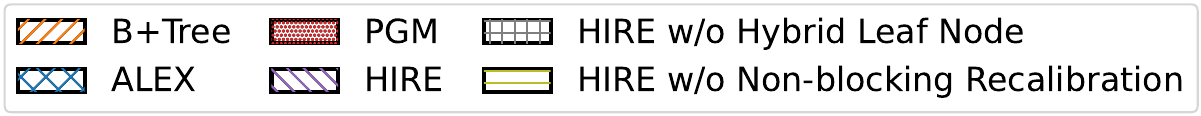}
\end{minipage}

\begin{minipage}{0.45\linewidth}
    \centering
    \includegraphics[width=0.85\columnwidth]{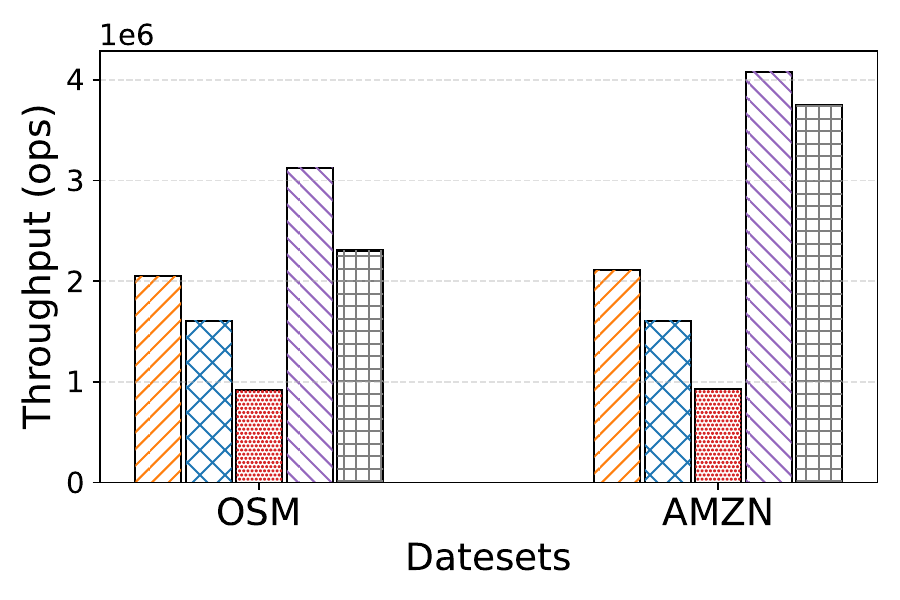}
\caption{Impact of Hybrid Nodes} 
	\label{fig:exp_hybrid_nodes}
\end{minipage}
\begin{minipage}{0.45\linewidth}
    \centering
	\includegraphics[width=0.85\columnwidth]{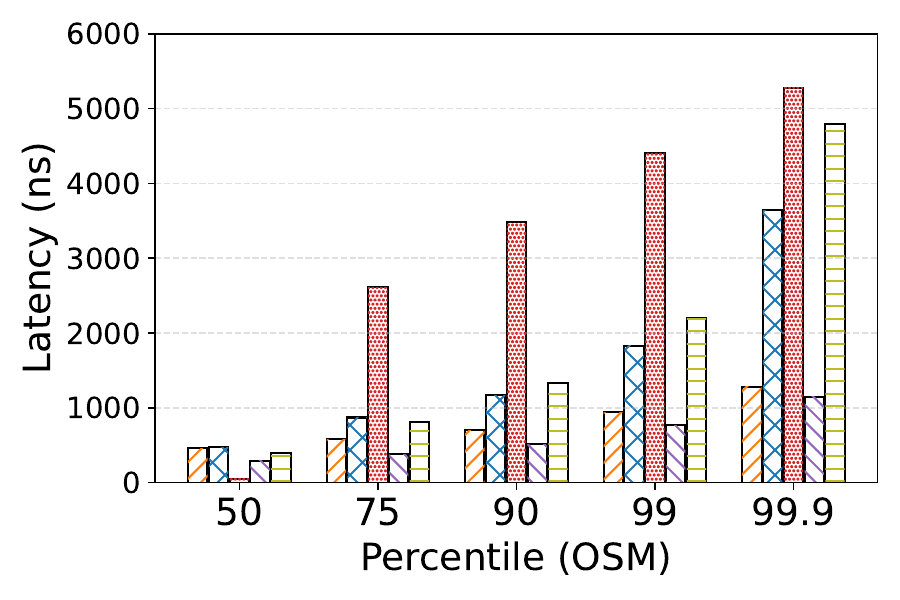}
    \caption{Impact of Non-blocking Recalibration} \label{fig:exp_recalibration}
\end{minipage}
\end{figure}

To evaluate our non-blocking, cost-driven recalibration design, we study a modified \opttree{} variant which replaces our proposed mechanism with a simplified, single-threaded, and blocking recalibration process. 
The results on the \textsf{OSM} dataset, shown in \cref{fig:exp_recalibration}, indicate that without our non-blocking recalibration design, the \opttree{} variant exhibits significant latency spikes. Tail latency at the 99th and 99.9th percentiles increases by $2.9\times$ and $4.2\times$, respectively, compared to the full \opttree{}.
These spikes result from large-scale structural adjustments, such as when a large model-based leaf node splits into multiple leaf nodes, forcing the index to stall for an extended period for structure modifications. This experiment confirms that our novel non-blocking recalibration mechanism is crucial for mitigating tail latency and ensuring robust performance, particularly under dynamic workloads.

\rone{\subsection{Performance on Concurrency} \label{sub_sec:exp_concurrency}
Although the initial design of \opttree{} did not support fully concurrent read-write operations, we observe that its structural modifications are localized, triggered only when legacy leaf nodes undergo splits, merges, or redistributions. This property allows us to develop a concurrent-safe version, denoted \opttree{}+, by incorporating a concurrency control mechanism similar to that of ALEX+ and LIPP+ in GRE~\cite{wongkham2022updatable}. This approach augmented the existing RCU mechanism in \opttree{} with node-level locking to ensure thread-safe access.
We evaluate the performance of \opttree{}+ under concurrent write-heavy workloads against other concurrent indexes, including B+-tree-OLC, ALEX+, and LIPP+. As PGM lacks a concurrent implementation, we replace it with FINEdex~\cite{li2021finedex}, which is natively designed for concurrency. All baseline implementations are sourced from the GRE framework~\cite{wongkham2022updatable}.

\begin{figure}
    \centering
    \begin{minipage}{0.76\linewidth}
        \centering
        \includegraphics[width=0.85\linewidth]{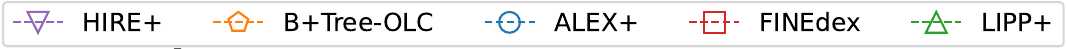}
    \end{minipage}
    \begin{minipage}{0.81\linewidth}
        \centering
        \includegraphics[width=0.9\linewidth]{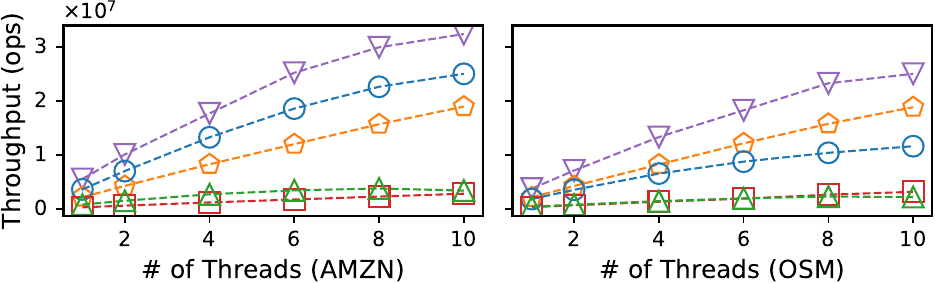}
    \end{minipage}
    \begin{minipage}{0.81\linewidth}
        \centering
        \includegraphics[width=0.9\linewidth]{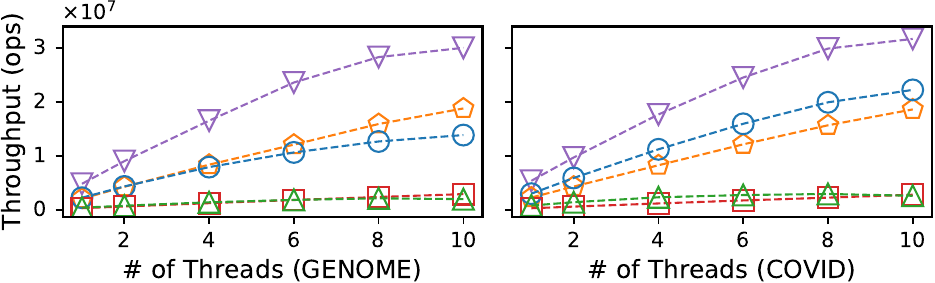}
    \end{minipage}
    \caption{\rone{Throughput Across Concurrent Threads (Write-Heavy)}}
    \label{fig:exp_concurrent}
    
\end{figure}

\cref{fig:exp_concurrent} illustrates the throughput of all indexes on four datasets as the number of concurrent threads scales from 1 to 10. \opttree{}+ demonstrates the best performance across all scenarios. \opttree{}+ achieves a $5.8$-$6.6\times$ speedup when concurrent threads scale from 1 to 10. At 10 threads, \opttree{}+ outperforms the best-performing baseline on each dataset: it is $1.3\times$ faster than ALEX+ on \textsf{AMZN}, $1.3\times$ faster than B+-tree-OLC on \textsf{OSM}, and $1.6\times$ and $1.4\times$ faster than the strongest baselines on \textsf{GENOME} and \textsf{COVID}, respectively.

We also observe the performance trade-offs of different concurrency control schemes. Indexes employing lock-free designs (e.g., B+-tree-OLC) or fine-grained locking (e.g., FINEdex) show sustained scalability. In contrast, other lock-based indexes, including our \opttree{}+, ALEX+, and LIPP+, hit a performance plateau beyond eight threads as lock contention escalates. This observation suggests that developing a more carefully designed lock-free concurrency mechanism for \opttree{} is a promising direction for future work.
}





    

\section{Related Work} \label{sec:related_works}

The concept of the learned index,  RMI~\cite{kraska2018case}, emerged in 2018 as a static approach. Early research on learned indexes mainly focuses on static workload. Several works, such as RadixSpline~\cite{kipf2020radixspline}, PLEX~\cite{stoian2021towards}, and Shift-Table~\cite{hadian2021shift}, demonstrate the potential of model-based indexing. However, a common criticism of these early designs is their inability to effectively support dynamic updates.



To overcome the limitation, updatable learned indexes have been introduced, including FITing-Tree~\cite{galakatos2019fiting}, PGM, and ALEX~\cite{ding2020alex}. Both FITing-Tree and PGM employ a balanced structure built bottom-up. ALEX, another notable updatable learned index, extends the RMI concept by constructing the index top-down. It sacrifices balance and a bounded structure to achieve better model fitting at different data granularities, and linearizes the data by adding gaps into segments to create a sparser structure. These pioneering updatable learned indexes delineate a design space characterized by one machine learning model (linear model), two construction methods (top-down and bottom-up), and three buffer types (index-level, segment-level, and data-level). Along with benchmarks like SOSD~\cite{kipf2019sosd} for performance evaluation on real-world data, these approaches have triggered a “Cambrian explosion” of learned index proposals.

During this surge, numerous learned indexes have been developed, revealing three main design trends. First, some works focus on optimizing model accuracy and structure for faster lookups. For example, LIPP achieves near error-free performance using a model-based insertion and splitting approach, although it sacrifices data locality. DILI~\cite{li2023dili} employs a concise, computation-efficient linear regression model for improved efficiency, while Hyper~\cite{zhang2024hyper} combines bottom-up and top-down construction approaches to balance latency and memory usage. SALI~\cite{ge2023sali} maintains internal statistical information to further refine its model. Second, other designs leverage modern hardware features to boost performance. For instance, Xindex~\cite{tang2020xindex} and FINEdex~\cite{li2021finedex} are designed to support concurrent operations. CARMI~\cite{zhang2021carmi} is the first learned index to fully exploit CPU cache characteristics. \cite{zhong2022learned} extends optimizations to GPUs.
\rone{
Third, a shift toward case-specific learned indexes has emerged to address the limitations of general-purpose designs. Most learned indexes, including our proposed \opttree{} and others like ALEX, PGM and LIPP, are designed for numeric data, as modeling the CDF of strings for key lookup is a non-trivial challenge. This has led to the development of specialized solutions. For string data, SIndex~\cite{wang2020sindex} and LITS~\cite{yang2024lits} represent notable recent progress. For other use cases, indexes like FLIRT~\cite{yang2023flirt} and SWIX~\cite{liang2024swix} have been designed specifically for indexing streaming windows.
}

There are also research efforts in optimizing learned indexes, including improving persistent memory efficiency~\cite{lu2021apex, cui2023learned, wang2024wipe, lan2024fully, lan2023simple, zhang2024making} and parameter tuning for optimal performance~\cite{chockchowwat2022tuning, chockchowwat2023airindex, wang2025new}.

\section{Conclusion} \label{sec:conclusion}
In this paper, we 
propose a novel hybrid learned index, \opttree{}, which delivers efficient and robust performance across diverse workloads and data distributions. Compared to state-of-the-art updatable learned indexes and traditional indexes, \opttree{} not only exhibits higher average performance but also significantly reduces tail latency. This approach greatly mitigates the risk of drastic performance drops under mixed workloads. Currently, \opttree{} provides a stable and efficient index for in-memory scenarios. Its balanced tree structure and compact leaf design offer a solid foundation for future expansion to disk-based environments, enabling support for larger-scale data and more complex workloads.

\begin{acks}
We are deeply thankful to the anonymous reviewers for their constructive reviews.
This work is supported by Hong Kong RGC Grants (Project No. 12202024 and 12200022) and the Swiss State Secretariat for Education, Research and Innovation (SERI) for the Prodasys project, a European Research Council (ERC) Advanced Grant.
Jianliang Xu is the corresponding author. 
\end{acks}

\bibliographystyle{ACM-Reference-Format}
\bibliography{ref}

\received{July 2025}
\received[revised]{October 2025}
\received[accepted]{November 2025}


\end{document}
\endinput